\documentclass[aps,superscriptaddress,nofootinbib,eqsecnum,prd,notitlepage,twocolumn]{revtex4-1} 

\usepackage{amssymb}   
\usepackage{amsmath}   
\usepackage{graphicx}  
\usepackage{graphicx}
\usepackage{subfigure}
\usepackage{hyperref}
\usepackage{float}
\usepackage{color}
\usepackage{slashed}
\usepackage{multirow}

\begin{document}

\title{Optimized perturbation theory applied to the study of
the thermodynamics and BEC-BCS crossover 
in the three-color Nambu--Jona-Lasinio model}

\author{Dyana C. Duarte}
\affiliation{ Departamento de F\'isica, Universidade Federal de Santa Maria, 
97105-900
Santa Maria, RS, Brazil}

\author{R. L. S. Farias}
\affiliation{ Departamento de F\'isica, Universidade Federal de Santa Maria, 
97105-900
Santa Maria, RS, Brazil}

\author{Pedro H. A. Manso}
\affiliation{Centro Federal de Educa\c{c}\~ao Tecnol\'ogica Celso Suckow da 
Fonseca,
Campus Maria da Gra\c{c}a, Rua Miguel \^{A}ngelo 96, 20785-223 Rio de Janeiro, 
RJ,
Brazil}

\author{Rudnei O.  Ramos}

\affiliation{Departamento de F\'{\i}sica Te\'orica, Universidade do
  Estado do Rio de Janeiro, 20550-013 Rio de Janeiro, RJ, Brazil}
\affiliation{Physics Department, McGill University, Montreal, Quebec, H3A 2T8, Canada}

\begin{abstract}

The Nambu--Jona-Lasinio model with two flavors, three colors and
diquark interactions is analyzed in the context of optimized
perturbation theory (OPT).  Corrections to the thermodynamical
potential that go beyond the large-$N_c$ (LN) approximation are
taken into account, and the  region of the phase diagram corresponding
to intermediate chemical potentials and very low temperatures is
explored. The  simultaneous presence of both the quark-antiquark
and diquark condensates can cause the
system to behave as a fluid composed of a Bose-Einstein condensate
(BEC) or a color superconductor one, in the form of a
Bardeen-Cooper-Schrieffer (BCS) superfluid. The
BEC-BCS crossover is then studied in the nonperturbative OPT scheme.  The
results obtained in the context of the OPT method are then contrasted
with those obtained in the LN approximation.  We show that there
are values for the coupling constants related  to quark-quark and
quark-antiquark interactions where the corrections beyond LN  brought
by the OPT method can influence the behavior of the diquark condensate
and the effective quark mass as a function of the baryon chemical
potential. These changes in the behavior of the phase structure of the
model modify the location of the critical  point related to the
phase structure as a whole of the model.  Also, when we impose the
color neutrality condition, our results show that the  nature of the
phase transition can change as well, shifting the ratio of the
quark-antiquark and quark-quark interactions to higher values in the
OPT case as compared to the LN approximation.

\end{abstract}

\maketitle

%%%%%%%%%%%%%%%%%%%%%%%%%%%%%%%%%%%%%%%%%%%%%%%%%%%%%%%%%%%%%%%%%%%%%%%%%%
\section{Introduction}
\label{intro}

Unveiling the phase structure of quantum chromodynamics (QCD) is
one active research area today. This is not only because of its
intrinsic theoretical interest, but also due to interest across many
different fields, ranging from the current heavy-ion collision 
experiments, to processes able to happen in the astrophysics of 
compact stellar objects like neutron stars, and also in cosmology.  
While QCD itself might be considered a well-defined theory, to study its properties
deep in the strong coupled nonperturbative regime, like at low
temperatures (energies) is notably extremely difficult. {}Furthermore,
when one also tries to study processes at high quark densities (large
chemical potential $\mu$), even state-of-the-art numerical techniques,
like lattice Monte Carlo QCD simulations (for a recent review, see,
e.g., Ref.~\cite{Ding:2015ona} and references therein), one faces
tremendous difficulties due to the so-called  ``sign problem''
(associated with the calculation of the  determinant of the quarks
matrix, which takes on a complex value when $\mu\neq 0$), and
progress in this direction has been painfully
slow~\cite{Cristoforetti:2012su,Fujii:2013sra}. As an alternative to
bypass the above mentioned difficulties, one typically recourses to low-energy
effective models for quantum chromodynamics (QCD), like for example
the Nambu--Jona-Lasinio (NJL) type of models~\cite{KLEVAN,BUBA},  
which are
valuable tools widely used to try to understand the underlying phase structure
of QCD, otherwise unaccessible either through the direct QCD Lagrangian
density or lattice QCD techniques. 

Of particular interest is the region of the QCD phase diagram at low
temperatures and  intermediate chemical potentials, even though  there
is still no consensus on the exact phase in which the quark matter is
expected to be found in this region. It corresponds to a portion of
the phase diagram not able to be probed by standard methods in QCD,
lattice QCD, or through current experiments on particle accelerators.
{}From the point of view of astrophysics, it is estimated that in
the cores of so-called  compact stars, these conditions are
present~\cite{SCHIMITT}. This then strongly motivates  studies towards
the understanding of the physics in this  region of the phase diagram,
sometimes also called the region of dense and cold matter,  through
the use of effective low-energy models that contain characteristics
common to  QCD, and which highlight some of the expected and relevant
behaviors for the system in that region. 

One of the most exciting possibilities is the occurrence of quark
Cooper pairing (color superconductivity) in this region of cold and
dense quark matter, a possibility that has been considered already 
in Ref.~\cite{Collins:1974ky}, and whose idea has gained
considerable interest since then (see, e.g., Ref.~\cite{Alford:2007xm}
for a detailed review on this subject).  In addition, many works have
considered the possibility the transition at low
temperature and baryon densities, going from the chiral broken phase 
to a color superconducting phase at large densities, could proceed through an
intermediary phase. In this intermediate phase, the quark matter  would undergo a crossover
between a regime where diquark pairs form difermion  molecules, giving
origin then to a Bose-Einstein condensation (BEC), and a weakly
coupled Bardeen-Cooper-Schrieffer (BCS) superfluid
phase~\cite{Abuki:2001be,HUANG1,RATTI,EBERTKLIM,CHATTER,NISHABUK,He:2007kd,ABUKBRAU,He:2010nb}
(for a recent review, see, e.g., Ref.~\cite{He:2013gga}).

Typically, we can employ an extended NJL model, where besides the
usual quark-antiquark four-Fermi interaction, which is responsible for
the formation of the chiral condensate of quark-antiquark pairs, a four-Fermi 
interaction for quark-quark, making possible
the formation of diquark condensates, akin to the pairing mechanism in
the BCS theory, as the magnitude of this coupling is increased. We can
then study the combined competition between these two types of
condensates in the system, the chiral and diquark ones. Several works
(see, e.g., Refs.~\cite{HUANG1,RATTI,EBERTKLIM,He:2010nb}) show that it is
possible for the condensate of diquarks initially to form a BEC phase,
before the system goes to the BCS state, as we increase the baryon
chemical potential going through this BEC-BCS 
crossover~\cite{NISHABUK,ABUKBRAU,CHATTER}.  Some possible observational
signatures for the BCS regime and the  possibility of coexistence
of the chiral and diquark phases  have been explored in the
literature~\cite{Kitazawa:2002bc},  while a connection with
high-temperature  superconductors and a possible pseudogap was studied
in  Refs.~\cite{Kitazawa:2001ft,Kitazawa:2003cs}.

We should note that the NJL model, since it does not include  gluon
degrees of freedom and thus cannot be used to study confinement, finds
applications in the low-energy (temperature) regime  of QCD and
quark matter, where the gluon degrees of freedom and their effects,
e.g., in the physics of (de)confinement, become less relevant.  But
this low-energy regime corresponds exactly to the regime where the
strong coupling and nonperturbative nature of the  nuclear matter is
relevant, which then likewise requires the use of appropriate nonperturbative
methods. The use of the NJL and similar models, as far as the
studies in this context are concerned, have mainly focused on the use
of the large-$N_c$ (LN) method (where $N_c$ is the number of colors),
or the Hartree approximation~\cite{BUBA}. In practice, the LN method
consists of making the change $G \rightarrow \lambda/(2N_{c})$ in the
four-Fermi quark-antiquark coupling constant of the model, by keeping 
$\lambda$ fixed
while making $N_{c}$ large and then keeping only the leading term in
the expansion when taking $N_c\to \infty$, even though we take $N_c=3$
at the very end for practical calculations.  However, such a method
cannot predict physical phenomena that might eventually  be related to
terms of the next order in $1/N_c$ in the expansion, which
have to be analyzed through some other self-consistent method and that
are required if we want to improve the precision of the results.  Even
though other approaches have been employed to obtain the thermodynamic
potential going beyond the leading LN result to study the phase 
structure of QCD  in the context of the
NJL model, these other methods can
quickly become more  involved or add further free parameters in the
analysis (let us recall that with the NJL model, being a nonrenormalizable
model, introducing higher-order corrections to the leading-order
thermodynamic potential is usually accompanied by the addition of
extra renormalization parameters), which is not always welcome.

In this work, we will make use of the OPT method (for a long, but still
far from complete, list of past applications of the OPT method in
quantum field theory problems, see e.g.,
Refs.~\cite{opt1,Stevenson:1981vj,Okopinska:1987hp,opt2,opt4,opt5,opt6,Pinto:1999py,Pinto:1999pg,FARIASOPT,opt8}).
In particular, let us mention that the OPT method has very
successfully been applied to NJL-similar types of models in low dimensions, in
particular in the  Gross-Neveu models in
2+1 dimensions~\cite{Kneur:2006ht}, revealing novel properties in
the phase diagram in that context e.g., a tricritical point, not
accessed by previous methods. Recent work on the OPT method tries to
combine its properties also with those of the renormalization group to
further push its applicability as far as renormalization  properties are
concerned~\cite{Kneur:2013coa,Kneur:2015moa,Kneur:2015uha}.

Previous applications of the OPT method for the study of the phase
structure in effective models of QCD include, for example, its use in
Walecka-type models~\cite{Krein:1995rp}, in the linear sigma
model~\cite{Khan:2016exa,opt10,opt11},  and also in the $SU(2)$ NJL
model~\cite{RUDOPT}, whose work in particular we will follow here closely, but in
the context of the NJL model with diquark interactions.  
As already mentioned, with the use of the NJL model, we can only hope to
capture some of the low-energy features of QCD at a qualitative level.
Given its nonrenormalizability and the other shortcomings already mentioned
above, the model itself is not a controlled approximation to QCD.
Yet, it is still a valuable tool, in particular to test different methods
that can improve over the simpler approximations used in the literature.
This is in particular true when considering the NJL in the context of 
the OPT approximation.
By going beyond the simple mean field theory, or LN approximation, the OPT
at first order in its implementation 
already includes some relevant mesonic fluctuations and,
in the present work, also contributions from the diquark interaction, which are
absent in the LN approximation and which would appear only at the next-to-leading
order in an expansion in $1/N_c$. In a way, we hope that by including these additional contributions
lacking in the LN case, we can improve the applicability of the NJL.
At the same time, we can also determine how the inclusion of these corrections
performs as compared to the LN approximation and determine 
whether they can provide both qualitatively and
quantitatively relevant corrections beyond the LN case that can be relevant
for QCD.
The aim of the present work is to present a detailed understanding of the
BEC-BCS crossover, making use of the nonperturbative OPT method  
and applying it to the NJL model
endowed with diquark interactions. We will analyze both the cases of
absence and presence of color neutrality, and we will use parameters
such that a comparison with previous results obtained within the LN
method in particular, those obtained by the authors of
Ref.~\cite{SUNHE} can be made.

This work is organized as follows: In Sec.~\ref{sec2}, we briefly
introduce the NJL model with diquark interactions.  In
Sec.~\ref{sec3}, we explain the OPT scheme and how it is applied to the present.
In Sec.~\ref{sec4}, we present the derivation of
the effective potential for the model and the relevant equations.  In
Sec.~\ref{sec5}, we discuss the determination of the parameters of the
model and the modifications required when applying the OPT scheme.  In
Sec.~\ref{results}, we perform our numerical analysis of the BEC-BCS
crossover, and the results obtained in the context of the OPT are
contrasted with those obtained in the LN approximation. In
Sec.~\ref{conclusions}, we present our conclusions. An Appendix is
included showing some of the technical details required in the
determination of the model parameters.

%%%%%%%%%%%%%%%%%%%%%%%%%%%%%%%%%%%%%%%%%%%%%%%%%%%%
\section{The NJL with diquark interactions} 
\label{sec2}

In this work, we will consider the NJL model with two flavors and three
colors ($N_c=3$) that includes both the usual chiral four-Fermi
quark-antiquark interaction and also the diquark channel, with the
Lagrangian density then given by~\cite{SUNHE,Kitazawa:2007zs,HUANG2,EBERTKLIM2,EBERTKLIM3}

\begin{eqnarray}
\mathcal{L}&=&
\bar{\psi}\left(i\gamma^{\mu}\partial_{\mu}-m\right)\psi  \nonumber
\\ &+& G_{s}\left[\left(\bar{\psi}\psi\right)^{2}
  +\left(\bar{\psi}i\gamma^{5}\vec{\tau}\psi\right)^2\right]
\nonumber \\ &+& \!\!\!\!\!\!\sum_{a=2,5,7} \!\!\!\!
G_{d}\left[\left(\bar{\psi}i\gamma^{5}
  \tau_{2}\lambda_{a}C\bar{\psi}^{T}\right)\left(\psi^{T}i\gamma^{5}\tau_{2}
  \lambda_{a}C\psi\right)\right],
\label{1LagNc3}
\end{eqnarray}
where $\psi$ represents the quark fields with a flavor doublet $(u,d)$
and color triplet ($N_{c}=3$), as well as a four-component Dirac
spinor. In Eq.~(\ref{1LagNc3}), $\vec \tau = (\tau_1,\tau_2,\tau_3)$
and $\lambda_a$ are the Pauli and Gell-Mann  matrices in the flavor and
color spaces, respectively.  $C\equiv i \gamma^{2}\gamma^{0}$ is the
charge conjugation operator, and $\psi^{T}$ is the transposed quark
field.  The mass $m$ 
is the current quark mass, while
$G_s$ and $G_d$ are the coupling constants for quark-antiquark and
quark-quark interactions, respectively.  In principle, these coupling
constants, if followed from  the QCD one-gluon exchange approximation
and from the {}Fierz
transformation~\cite{RATTI,HUANG2}, would be related like $G_{d}=N_{c}/(N_{c}-1)
G_{s}/2$, such that for $N_c=3$, then $G_d=3G_s/4$.  Here, however,
we follow the philosophy of Refs.~\cite{SUNHE,EBERTKLIM}, where these
couplings are treated as free parameters and  we will not fix
relations between them. In practice, as we will see later on when
studying the numerical results, there will always be a ratio of
couplings $G_d/G_s$ below a certain minimum value 
such that the transition from the chiral phase to the diquarks
with a nonvanishing vacuum expectation value will tend to be
first order, thus preventing a BEC phase, while for larger values of
the ratio there will be a maximum value for this ratio  beyond which
diquarks would already condense at a baryonic chemical potential
$\mu_B=0$ i.e., the diquarks would become massless and the vacuum
unstable~\cite{EBERTKLIM3,ZHUANG}. We will
discuss theses issues in more detail later on in the text .

By making use of a Hubbard-Stratonovich transformation in
Eq.~(\ref{1LagNc3}),  the four-Fermi interactions can be rewritten in
terms of bosonic fields, given by $\zeta$,
$\vec{\pi}$, $\phi_{a}$, and $\phi^{*}_{a}$, and can be expressed in the form

\begin{widetext}
\begin{eqnarray}
 \mathcal{L}_{B}&=&\bar{\psi}\left[i\gamma^{\mu}\partial_{\mu}-m-(\zeta
   +i\gamma^5\vec{\pi}\cdot\vec{\tau})\right]\psi+\frac{1}{2}
\sum_{a=2,5,7}\psi^{T}\phi_{a}^*i
 \gamma^5\tau_{2}\lambda_{a}C\psi   \nonumber\\ &+&
 \frac{1}{2}\sum_{a=2,5,7}\bar{\psi}\phi_{a}
 i\gamma^5\tau_{2}\lambda_{a}C\bar{\psi}^T
 -\frac{1}{4G_s}\left(\zeta^2+\vec{\pi}^2\right)-\frac{1}{4G_d}
\sum_{a=2,5,7}|\phi_{a}|^2. 
\label{LB}
\end{eqnarray}
\end{widetext}
{}From the use of the Euler-Lagrange equations for $\zeta$,
$\vec{\pi}$, $\phi_{a}$, and $\phi^{*}_{a}$,  we have that

\begin{eqnarray} \label{condzetapi}
&&\zeta=-2G_{s}\bar{\psi}\psi,
  \\   &&\vec{\pi}=-2G_{s}\bar{\psi}i\gamma^{5}\vec{\tau}\psi,
  \\ &&\phi_{a}=2G_{d}\psi^{T}C i\gamma^{5}\tau_{2}\lambda_{a}\psi,
  \\   &&\phi^{*}_{a}=-2G_{d}\bar{\psi}
  i\gamma^{5}\tau_{2}\lambda_{a}C \bar{\psi}^{T},
\end{eqnarray}
and upon substitution in Eq.~(\ref{LB}), we recover the original
Lagrangian density of Eq.~(\ref{1LagNc3}).

We will consider, without loss of generality (see, for instance,
Ref.~\cite{SUNHE}),  that only the quarks with colors 1 and 2  form
diquarks. This condition is satisfied when $\phi_{2}=\phi$ and
$\phi_{5}=\phi_{7}=0$. Therefore, we can write the Lagrangian density
(\ref{LB}) in the form

\begin{widetext}
\begin{eqnarray}
\mathcal{L}_{B} &=&\bar{q}_{3}\left[i\gamma^{\mu}\partial_{\mu}-m
  -(\zeta+i\gamma^5\vec{\pi}\cdot\vec{\tau})\right]q_{3}+
\bar{q}_{1,2}\left[i\gamma^{\mu}
  \partial_{\mu}-m-(\zeta+i\gamma^5\vec{\pi}\cdot\vec{\tau})\right]q_{1,2}
\nonumber\\ &+&
\frac{\phi^{*}}{2}q_{1,2}^{T}iC\gamma^5\tau_{2}t_{2}q_{1,2} +
\frac{\phi}{2}\bar{q}_{1,2}
i\gamma^{5}C\tau_{2}t_{2}\bar{q}_{1,2}^{T}-
\frac{1}{4G_s}\left(\zeta^2+\vec{\pi}^2\right)
-\frac{1}{4G_d}|\phi|^2, \label{LBfinal}
\end{eqnarray}
\end{widetext}
where we have used

\begin{eqnarray} \label{defpsiT}
 \psi^{T} \equiv \begin{pmatrix} q_{1,2}^{T} & q_{3}^{T}\end{pmatrix}
 \qquad    q_{1,2}^{T} \equiv \begin{pmatrix} q_{1}^{T} & q_{2}^{T} 
\end{pmatrix},
\end{eqnarray}
with $q_{i}$ ($i=1,2,3$) representing the color quark fields and
$t_{2}$ being the second Pauli matrix in the color space 1 and 2
(green and red).  This form  of Eq.~(\ref{LBfinal}) makes explicit the
fact that quarks with color 3 (blue)  do not participate in the
formation of the diquark condensate.

%%%%%%%%%%%%%%%%%%%%%%%%%%%%%%%%%%%%%%%%%%%%%%%%%%%%%%%%%%%%%%%%%%%%%%%%%
\section{The OPT applied to the NJL model with diquarks}
\label{sec3}

The OPT method consists of initially defining an interpolated (or
deformed) Lagrangian density in the form

\begin{eqnarray}
\mathcal{L}_{\delta}&\equiv&\delta\mathcal{L}+(1-\delta)
\mathcal{L}_{0}(\eta_{i})
\nonumber\\ &=&\mathcal{L}_{0}(\eta_{i})+\delta
\left[\mathcal{L}-\mathcal{L}_{0}(\eta_{i})\right], 
\label{LdeltaOPT}
\end{eqnarray}
where $\mathcal{L}_{0}$ is the Lagrangian density of a solvable theory,
modified by  the introduction of arbitrary parameters (which in the
present model will be associated with the  interaction channels
between fermions and their condensates) with  mass dimension terms
$\eta_{i}$, and $\delta$ is a (bookkeeping) parameter that is  used to
enable a perturbative expansion; it is set to $\delta=1$ at the
end.  We can note that if $\delta=1$, we then immediately recover the
Lagrangian density of the original  theory; if $\delta=0$, we have the
solvable Lagrangian density $\mathcal{L}_{0}$.  Any physical quantity $\mathcal{P}_{n}$ that is calculated
up to a given order $n$ in  $\delta$ will, however, depend explicitly
on the parameters  $\eta_{i}$, which are not part of the original
theory. Thus, we must impose an appropriate  condition that best  fixes
the values for these arbitrary parameters in a self-consistent way.
The criterion we will use in this work, which was also used in many other
previous OPT applications (for other alternative optimization
criteria, see, e.g., Refs.~\cite{opt6,FARIASOPT}), is the principle of
minimum sensitivity (PMS), by requiring that~\cite{Stevenson:1981vj}

\begin{eqnarray} \label{defPMS}
\left. \frac{\partial \mathcal{P}_{n}}{\partial
  \eta_{i}}\right|_{\eta_{i}=\bar{\eta}_{i}}=0,
\end{eqnarray}
through which the parameters $\bar{\eta}_{i}$ are those that make the
computed quantities an extremum (a minimum) with respect to these mass
parameters and guarantee that $\mathcal{P}_{n}$ is  locally
independent (sensitive) of $\bar{\eta}_{i}$.  The convergence of the
OPT under different contexts has been  shown in the many papers 
cited in Ref.~\cite{opt6}.

The interpolation in the present model is performed as follows:
Starting from the Lagrangian density in terms of the auxiliary fields,
Eq.~(\ref{LB}), and following, e.g., the procedure shown in
Ref.~\cite{RUDOPT}, we can  define the OPT Lagrangian density
$\mathcal{L}_0$ in Eq.~(\ref{LdeltaOPT}) as

\begin{eqnarray}
\!\!\!\!\!\!\!\mathcal{L}_{0}&\equiv&
\bar{\psi}\left(i\gamma^{\mu}\partial_{\mu}-m\right)\psi
-\bar{\psi}(\eta+i\gamma^{5}\vec{\tau}\cdot \vec{\eta}_{\pi})\psi
\nonumber\\ &-&\!\!\!\!\!\!\!\sum_{a=2,5,7} \!\!\!\! \left(
\bar{\psi}i\gamma^{5}\tau_{2}\lambda_{a}C\alpha_{1a}'\bar{\psi}^{T}
\!+\!\psi^{T}i\gamma^{5}\tau_{2}\lambda_{a}C\alpha_{2a}'\psi \right). 
\label{Lparameters}
\end{eqnarray}
In Eq.~(\ref{Lparameters}), the OPT mass parameters $\eta$ and $\eta_{\pi}$ are
the ones related to the scalar and pseudoscalar channels,
respectively, while $\alpha_{1a}'$ and $\alpha_{2a}'$ are those for
the quark-quark interaction scalar channel.  Since $\langle
\vec{\pi} \rangle =0$,  we can then set $\vec{\eta}_{\pi}=0$
consistently~\cite{RUDOPT}. 

Overall, the interpolated Lagrangian density used in the OPT
scheme in the present model can then be expressed in the form

\begin{widetext}
\begin{eqnarray}
\mathcal{L}_{\delta,B}&=&-\delta\frac{(\zeta^{2}+\vec{\pi}^{2})}{4G_{s}}
-\delta\frac{|\phi|^{2}}{4G_{d}}+\bar{\psi}\left[i\gamma^{\mu}\partial_{\mu}
  -m-\delta\left(\zeta+i\gamma^{5}\vec{\tau}\cdot\vec{\pi}\right)
-(1-\delta)\eta\right]\psi
\nonumber\\ &+&\psi^{T}Ci\gamma^{5}\tau_{2}\frac{\lambda_{2}}{2}
\left[\delta\phi^{*}
  +(1-\delta)\alpha_{2}\right]\psi+\bar{\psi}i\gamma^{5}\tau_{2}
\frac{\lambda_{2}}{2}C
\left[\delta\phi+(1-\delta)\alpha_{1}\right]\bar{\psi}^{T} \ , 
\label{LBOPTestNJL}
\end{eqnarray}
\end{widetext}
in which we have defined $\alpha_{1}\equiv-2\alpha_{12}'$  and
$\alpha_{2}\equiv-2\alpha_{22}'$, and we have also again performed  the
rotation $\phi_{2}\equiv\phi$  and $\phi_{5}=\phi_{7}=0$, resulting in
$\alpha_{1k}=\alpha_{2k}=0$, with  $k=5,7$.  It is important to note
that when $\delta=1$, we retrieve the original theory given by
Eq.~(\ref{LB}).  We can also conveniently rewrite Eq.~(\ref{LBOPTestNJL})
as 

\begin{widetext}
\begin{eqnarray}
\mathcal{L}_{\delta,B} &=&
\bar{q}_{3}\left[i\gamma^{\mu}\partial_{\mu}-m-\delta(\zeta
  +i\gamma^5\vec{\pi}\cdot\vec{\tau})-(1-\delta)\eta \right]q_{3} +
\bar{q}_{1,2}
\left[i\gamma^{\mu}\partial_{\mu}-m-\delta(\zeta+i\gamma^5
\vec{\pi}\cdot\vec{\tau})
  -(1-\delta)\eta \right]q_{1,2} \nonumber\\ &+&
\frac{1}{2}q_{1,2}^{T}iC\gamma^5\tau_{2}t_{2}\left[\delta\phi^{*}+(1-\delta)
  \alpha_{2} \right]q_{1,2} +
\frac{1}{2}\bar{q}_{1,2}i\gamma^{5}C\tau_{2}t_{2}
\left[\delta\phi+(1-\delta)\alpha_{1} \right]\bar{q}_{1,2}^{T} -
\frac{\delta}{4G_s}\left(\zeta^2+\vec{\pi}^2\right)
-\frac{\delta}{4G_d}|\phi|^2.  \nonumber\\ \label{LBfinaldelta}
\end{eqnarray}
\end{widetext}

%%%%%%%%%%%%%%%%%%%%%%%%%%%%%%%%%%%%%%%%%%%%%%%%%%%%%%%%%%%%%%%%%%%%%%%%%
\section{The effective potential in the OPT method}
\label{sec4}

We are now in position to derive the thermodynamic effective potential
for  the NJL model with diquark interactions within the OPT scheme. We
evaluate the effective potential up to order $\delta^{1}$ in the OPT method,
which will by itself already supply us with correction terms going
beyond the standard LN approximation.

\begin{figure*}[htpb]
\includegraphics[width=12cm]{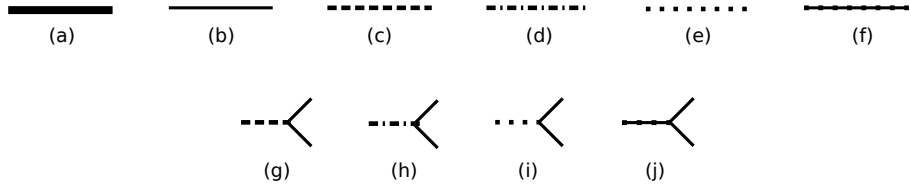}
\caption[fgt]{{\footnotesize The representation of the Feynman rule
    elements in the OPT method.  (a) Fermionic propagator related to
    the $\delta$-dependent quark $\Psi$-field.  (b) Fermionic
    propagator when $\delta=0$.  (c) Propagator related to the boson
    (auxiliary) $\zeta$-field.  (d) Propagator related to each of
    three $\pi_{i}$-fields.  (e) Propagator related to the diquark
    auxiliary field, real component $\phi_{R}$.  (f) Propagator
    related to the diquark auxiliary field, imaginary component
    $\phi_{I}$. (g), (h), (i), (j), The vertices related to
    the point interactions between a bosonic and two fermionic  fields
    at $\delta=0$. (If fermionic propagators depend on $\delta$, the lines
    are represented here as thick ones.)}}
\label{fig: FR}
\end{figure*}

All relevant {}Feynman rules regarding the propagators and vertices
within the OPT scheme are represented in {}Fig.~\ref{fig: FR}.

%%%%%%%%%%%%%%%%%%%%%%%%%%%%%%%%%%%%%%%%%%%%%%%%%%
%%%%%%%%%%%%%%%%%%%%%%%%%%%%%%%%%%%%%%%%%%%%%%%%%%%%

\begin{figure*}[!htb]
\includegraphics[width=10cm]{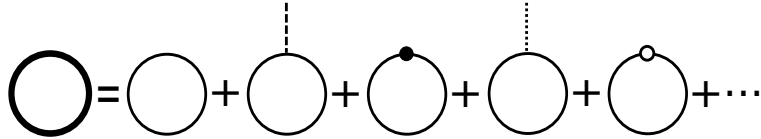}
\caption[dia]{{\footnotesize One-loop diagrams in OPT expanded up  to
    order $\delta^{1}$. The thick continuous line represents the fermionic
    propagator as a function  of $\delta$. Thin lines represent the
    propagator when $\delta=0$. The dashed line represents the
    $\sigma$-field propagator, while the dotted one is associated with
    the $\Delta$-field propagator.  The large black dot  represents a
    $\delta\eta$-vertex insertion, and the white one represents a
    $\delta\alpha$-vertex  insertion.}}
\label{fig: FD1Loop}
\end{figure*}

\begin{figure*}[!htp]
\includegraphics[width=12cm]{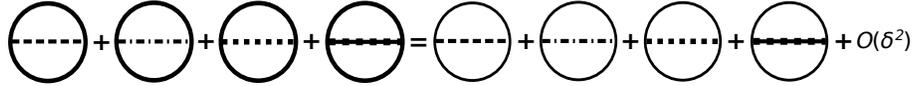}
\caption[edc]{{\footnotesize To the left of the equality, the sum of
    two-loop diagrams  in OPT ($\propto 1/N_{c}^{1}$) capable of
    generating contributions of order $\delta^{1}$.  To the right of the 
    equality, sum of two-loops diagrams when expanded up to order $\delta^1$. All
    elements are defined according to the legend of Fig.~\ref{fig:
      FR}.}}
\label{fig: FD2Loops}
\end{figure*}

Up to order $\delta$ in the OPT, we will have both one-loop terms, as
shown in {}Fig.~\ref{fig: FD1Loop}, and also two-loop terms, as shown in
{}Fig.~\ref{fig: FD2Loops}.

%%%%%%%%%%%%%%%%%%%%%%%%%%%%%%%%%%%%%%%%%%%%%%%%%%%%
%%%%%%%%%%%%%%%%%%%%%%%%%%%%%%%%%%%%%%%%%%%%%%%%%%%

%%%%%%%%%%%%%%%%%%%%%%%%%%%%%%%%%%%%%%%%%%%%%%%%%%%%%
%%%%%%%%%%%%%%%%%%%%%%%%%%%%%%%%%%%%%%%%%%%%%%%%%%%%%%

Below, we will evaluate separately the one-loop and two-loop contributions
shown in {}Figs.~\ref{fig: FD1Loop} and \ref{fig: FD2Loops}, respectively.

%%%%%%%%%%%%%%%%%%%%%%%%%%%%%%%%%%%%%%%%%%%%%%%%%%%%%%%%%%%%%%%%%%%
\subsection{The one-loop contribution to the effective potential 
in the OPT expansion} 

At order $\delta$, the one-loop {}Feynman' diagrams contribution to
the effective potential in the OPT is shown in {}Fig.~\ref{fig:
  FD1Loop}.  In {}Fig.~\ref{fig: FD1Loop}, a full line is associated
with a fermionic propagator related to all quarks, and it is a function
of $\delta$.  By expanding it in powers of $\delta$ and truncating to
${\cal O}(\delta^{1})$,  we obtain the resulting contributions shown
on the right-hand side of {}Fig.~\ref{fig: FD1Loop}. 
The effective potential can be obtained by the usual functional 
integral technique most easily when one makes use of
the standard Nambu-Gor'kov formalism~\cite{gorkov,Nambu:1960tm}
applied to the quark fields.
By also using the
Matsubara formalism of finite-temperature quantum field
theory~\cite{kapusta} and performing the sum over the Matsubara
frequencies for the fermions, the obtained effective potential at
one loop and order $\delta$ at  finite temperature ($T=1/\beta$) and
chemical potential  can be expressed explicitly in the form

\begin{widetext}
\begin{eqnarray}
V_{{\rm eff};1-{\rm loop},
  \delta^1}^{\mu,\mu_{b},\beta}&=&\delta\frac{\sigma^{2}}{4G_{s}}+
\delta\frac{|\Delta|^{2}}{4G_{d}}
\nonumber\\ &-&{\cal N} \int{\frac{d^{3}p}{(2\pi)^{3}}\left\{\left[E_{\alpha
      \vec{p}}^{-}(\mu) +E_{\alpha
      \vec{p}}^{+}(\mu)\right]+\delta\left[\frac{d}{d\delta}E_{\alpha
      \vec{p}}^{\delta-}(\mu) +\frac{d}{d\delta}E_{\alpha
      \vec{p}}^{\delta+}(\mu)\right]_{\delta=0}\right\}}
\nonumber\\ &-&N_{f}\int
\frac{d^{3}p}{(2\pi)^{3}}\left\{\left[E_{\vec{p}}^{-}(\mu_{b})
+E_{\vec{p}}^{+}(\mu_{b})\right]
+\delta\left[\frac{d}{d\delta}E_{\vec{p}}^{\delta-}(\mu_{b})
  +\frac{d}{d\delta}E_{\vec{p}}^{\delta+}(\mu_{b})\right]_{\delta=0}\right\}
\nonumber\\ &-& 2{\cal N}\int
\frac{d^{3}p}{(2\pi)^{3}}\left\{\frac{1}{\beta}\ln{\left(1 +e^{-\beta
    E_{\alpha \vec{p}}^{-}(\mu)}\right)}+\frac{1}{\beta}\ln{\left(1
  +e^{-\beta E_{\alpha \vec{p}}^{+}(\mu)}\right)}
\right. \nonumber\\ &-&\left.\delta\left[\frac{1}{e^{\beta E_{\alpha
        \vec{p}}^{\delta-}(\mu)} +1}\frac{d}{d\delta}E_{\alpha
    \vec{p}}^{\delta-}(\mu)+\frac{1}{e^{\beta E_{\alpha
        \vec{p}}^{\delta+}(\mu)} +1}\frac{d}{d\delta}E_{\alpha
    \vec{p}}^{\delta+}(\mu)\right]_{\delta=0}\right\}
\nonumber\\ &-&2N_{f}\int
\frac{d^{3}p}{(2\pi)^{3}}\left\{\frac{1}{\beta}\ln{\left(1 +e^{-\beta
    E_{\vec{p}}^{-}(\mu_{b})}\right)}+\frac{1}{\beta}\ln{\left(1 +e^{-\beta
    E_{\vec{p}}^{+}(\mu_{b})}\right)}
\right. \nonumber\\ &-&\left.\delta\left[\frac{1}{e^{\beta
      E_{\vec{p}}^{\delta-}(\mu_{b})}
    +1}\frac{d}{d\delta}E_{\vec{p}}^{\delta-}(\mu_{b})+\frac{1}{e^{\beta
      E_{\vec{p}}^{\delta+}(\mu_{b})}
    +1}\frac{d}{d\delta}E_{\vec{p}}^{\delta+}(\mu_{b})\right]_{\delta=0}\right\}, 
\label{VeffMFOPT}
\end{eqnarray}
\end{widetext}
where we have defined ${\cal N}\equiv (N_{c}-1)N_{f}$ and

\begin{eqnarray}
&&\!\!\!E_{\alpha
    \vec{p}}^{\delta\pm}(\mu')\!=\!\sqrt{[E_{\vec{p}}^{\delta\pm}(\mu')]^{2}
    +\prod_{s=1}^{2}[\alpha_{s}+\delta(\Delta_{s}-\alpha_{s})]} ,
\label{fermion-E}
\\ &&E_{\vec{p}}^{\delta\pm}(\mu')=E_{\vec{p}}^{\delta} \pm \mu',
\\ &&E_{\vec{p}}^{\delta}=\sqrt{\vec{p}^{2}+[M'_{\eta}(\delta,\sigma,0)]^{2}},
\label{dispOrig}
\end{eqnarray}
with $\mu'= \mu,\mu_{b}$ and

\begin{eqnarray}
M'_{\eta}(\delta,\sigma,0)&\equiv&
M'_{\eta}(\delta,\langle\zeta\rangle,\langle\vec{\pi}\rangle)
\nonumber\\ &=&m+\delta(\langle\zeta\rangle+i\gamma^5
\langle\vec{\pi}\rangle\cdot\vec{\tau})
+(1-\delta)\eta
\nonumber\\ 
&=&m+\eta+\delta(\sigma-\eta) \label{M'etadelta}, 
\end{eqnarray}
where we have used $\langle \zeta \rangle = \sigma$ and $\langle
\vec{\pi} \rangle = 0$.  Note that the chemical  potentials were
included by the usual prescription  $i\partial_{0} \rightarrow
i\partial_{0} + \mu_j$   in Eq.~(\ref{LBfinaldelta}), with
$\mu_{1}\equiv\mu_{r}$, $\mu_{2}\equiv \mu_{g}$ and  $\mu_{3}\equiv
\mu_{b}$, but in order to ensure that $SU(2)$ color symmetry between red and green quarks
is not explicitly broken, we take $\mu_{g}=\mu_{r}\equiv \mu$.
In addition, we have that
$\Delta_{1}\equiv \Delta$, $\Delta_{2}\equiv \Delta^{*}$,  with
$\langle \phi \rangle = \Delta$ and $\langle \phi^* \rangle =
\Delta^*$. (Without loss of generality, we will assume
$\Delta^{*}=\Delta$ and  $\alpha_{1}=\alpha_{2}\equiv \alpha$, since
only the absolute values of these quantities appear at the end.)

Then, in Eq.~(\ref{VeffMFOPT}), we have that

\begin{eqnarray}
\left.\frac{d}{d\delta}E_{\vec{p}}^{\delta\pm}(\mu')\right|_{\delta=0}
&=&\frac{(m+\eta)(\sigma-\eta)}{E_{\vec{p}}},
\label{derivative1}
\\ \left.\frac{d}{d\delta}E_{\alpha\vec{p}}^{\delta\pm}(\mu')\right|_{\delta=0}
&=&\frac{1}{E_{\alpha\vec{p}}^{\pm}(\mu')}
\left[\frac{E_{\vec{p}}^{\pm}(\mu')}{E_{\vec{p}}}(m+\eta)(\sigma-\eta)
  \right.\nonumber\\ &+&\left.\alpha(\Delta-\alpha)
  \vphantom{\frac{E_{\vec{p}}^{i\pm}(0)}{E_{\vec{p}}(0)}}\right],
\label{derivative2}
\end{eqnarray}
with

\begin{eqnarray}
&&E_{\alpha
    \vec{p}}^{\pm}(\mu')\equiv E_{\alpha\vec{p}}^{\delta=0\pm}(\mu')
=\sqrt{[E_{\vec{p}}^{\pm}(\mu')]^{2}+\alpha^{2}},
  \nonumber\\ &&E_{\vec{p}}^{\pm}(\mu')\equiv E_{\vec{p}}^{\delta=0\pm}(\mu')= E_{\vec{p}} \pm \mu',
  \nonumber\\ &&E_{\vec{p}}\equiv E_{\vec{p}}^{\delta=0}=\sqrt{\vec{p}^{2}+(m+\eta)^{2}}.
\label{EnergiesOPT}
\end{eqnarray}

Note that if we were already at this one-loop level of the OPT
expansion, to apply the PMS condition Eq.~(\ref{defPMS}) to
Eq.~(\ref{VeffMFOPT}) to determine the optima  $\bar{\eta}$ and
$\bar{\alpha}$, obtained, respectively, from

\begin{eqnarray} 
\frac{\partial V_{\rm eff}}{\partial \eta}\Bigr|_{\eta=\bar{\eta}}
=0,
\label{pmseta}
\end{eqnarray}
and

\begin{equation}
\frac{\partial V_{\rm eff}}{\partial
  \alpha}\Bigr|_ {\alpha=\bar{\alpha}}=0, 
\label{pmsalpha}
\end{equation}
we would simply find that $\bar{\eta}=\sigma$ and $\bar{\alpha}=\Delta$,
thus recovering immediately the usual LN approximation [note that in
this case, the terms in Eq.~(\ref{VeffMFOPT}) involving the
derivatives (\ref{derivative1}) and (\ref{derivative2}) vanish], as
expected.

%%%%%%%%%%%%%%%%%%%%%%%%%%%%%%%%%%%%%%%%%%%%%%%%%%%%%%%%
\subsection{The contributions of two loops in the OPT expansion}

Let us now give the explicit expressions for the two-loop diagrams
that also contribute with terms of order  $\delta^{1}$ in the OPT
expansion.  At two loops the diagrams that contribute at order
$\delta^{1}$ are those shown in  {}Fig.~\ref{fig: FD2Loops}, which
are constructed from the Feynman rules where, 
from the Lagrangian density in the OPT interpolation,
Eq.~(\ref{LBfinaldelta}), we have that the fermionic propagators carry
a dispersion relation dependent on $\delta$, as given by Eq.~(\ref{fermion-E});
each (nonpropagating) bosonic propagator contributes with a factor
$\delta^{-1}$; and each interaction vertex carries a factor
$\delta$. It is useful to separate the contributions that contribute
explicitly on the diquark OPT mass parameter $\alpha$,  corresponding
to the contributions that involve the green and red quarks, and the
ones involving the blue quark (which has $\alpha=0$).

The two-loop contributions of order $\delta$ from the OPT expansion,
and the contributions to the effective potential  at finite
temperature and chemical potential due to the quarks with colors 1
(red) and 2  (green), which are the ones forming diquarks, can be
expressed explicitly in the form

\begin{eqnarray}
\lefteqn{ \!\!\!\!\!\!\!\!\!\!  \tilde{V}_{{\rm eff}; 2-{\rm
      loop},\delta^1}^{\mu,\beta} =
  {\cal N}\delta\left[(n_{\pi}+1)G_{s}
    -2G_{d}\right]\left[{\cal F}_1(\eta,\alpha,\mu,T) \right]^{2}  }
\nonumber\\ &&\!\!\!\!\!\!\!\!\!-
{\cal N}\delta(n_{\pi}+1)G_{s}\left[{\cal F}_2(\eta,\alpha,\mu,T) \right]^{2}
\nonumber\\ &&\!\!\!\!\!\!\!\!\! -{\cal N}\delta
(m+\eta)^{2}\left[(n_{\pi}-1)G_{s}+2G_{d}\right]
\left[{\cal F}_3(\eta,\alpha,\mu,T) \right]^{2}
\nonumber\\ &&\!\!\!\!\!\!\!\!\!+
{\cal N}\delta (m+\eta)^{2}(n_{\pi}-1)G_{s}
\left[{\cal F}_4(\eta,\alpha,\mu,T) \right]^{2}, \label{Veff2loopq12}
\end{eqnarray}
where again we are using ${\cal N}=(N_{c}-1)N_{f}$, $n_{\pi}=3$ is the
number of pions, and we have defined the functions

\begin{eqnarray}
{\cal F}_1(\eta,\alpha,\mu,T) &=&
\int\frac{d^{3}p}{(2\pi)^{3}}
    \left[\frac{E_{\vec{p}}^{+}(\mu)}{E_{\alpha\vec{p}}^{+}(\mu)}
    \left(\frac{1}{2}-\frac{1}{e^{\beta E_{\alpha
          \vec{p}}^{+}(\mu)}+1}\right)\right.
\nonumber\\
&-& \left. \frac{E_{\vec{p}}^{-}(\mu)}{E_{\alpha\vec{p}}^{-}(\mu)}
    \left(\frac{1}{2}-\frac{1}{e^{\beta E_{\alpha \vec{p}}^{-}(\mu)}
      +1}\right)\right],
\end{eqnarray}

\begin{eqnarray}
{\cal F}_2(\eta,\alpha,\mu,T) &=&
\int\frac{d^{3}p}{(2\pi)^{3}}
  \left[\frac{\alpha}{E_{\alpha\vec{p}}^{+}(\mu)}\left(\frac{1}{2}
  -\frac{1}{e^{\beta E_{\alpha \vec{p}}^{+}(\mu)}+1}\right) \right.
\nonumber \\
&+& \left.\frac{\alpha}{E_{\alpha\vec{p}}^{-}(\mu)}
  \left(\frac{1}{2}-\frac{1}{e^{\beta E_{\alpha \vec{p}}^{-}(\mu)}
    +1}\right)\right],
\end{eqnarray}

\begin{eqnarray}
{\cal F}_3(\eta,\alpha,\mu,T) &=&
\int\frac{d^{3}p}{(2\pi)^{3}E_{\vec{p}}}
  \left[\frac{E_{\vec{p}}^{+}(\mu)}{E_{\alpha\vec{p}}^{+}(\mu)}\left(\frac{1}{2}
  -\frac{1}{e^{\beta E_{\alpha \vec{p}}^{+}(\mu)}+1}\right) \right.
\nonumber\\
&+& \left. \frac{E_{\vec{p}}^{-}(\mu)}{E_{\alpha\vec{p}}^{-}(\mu)}
  \left(\frac{1}{2}-\frac{1}{e^{\beta E_{\alpha \vec{p}}^{-}(\mu)}
    +1}\right)\right],
\end{eqnarray}

\begin{eqnarray}
{\cal F}_4(\eta,\alpha,\mu,T) &=&
\int\frac{d^{3}p}{(2\pi)^{3}E_{\vec{p}}}
  \left[\frac{\alpha}{E_{\alpha\vec{p}}^{+}(\mu)}\left(\frac{1}{2}
  -\frac{1}{e^{\beta E_{\alpha \vec{p}}^{+}(\mu)}+1}\right) \right.
\nonumber\\
&-& \left. \frac{\alpha}{E_{\alpha\vec{p}}^{-}(\mu)}
  \left(\frac{1}{2}-\frac{1}{e^{\beta E_{\alpha
        \vec{p}}^{-}(\mu)}+1}\right) \right].
\end{eqnarray}

The two-loop terms' contributions to the effective potential  for quarks
with color 3 (blue) are very analogous to the ones derived in
Ref.~\cite{RUDOPT}, and they can be obtained directly from
Eq.~(\ref{Veff2loopq12}) by simply making the changes $\alpha\rightarrow 0$, ${\cal N}
\rightarrow  N_{f}$, $G_{d}\rightarrow 0$,  $\mu\rightarrow
\mu_{b}$, and $[E_{\alpha\vec{p}}^{\pm}(\mu),E_{\vec{p}}^{\pm}(\mu)]
\rightarrow [E_{\vec{p}}^{\pm}(\mu_{b}),E_{\vec{p}}^{\pm}(\mu_{b})]$, from which
we obtain

\begin{widetext}
\begin{eqnarray}
\bar{V}_{{\rm eff};2-{\rm
    loop},\delta^1}^{\mu_{b},\beta}&=&N_{f}\delta(n_{\pi}+1)G_{s}
\left[\int\frac{d^{3}p}{(2\pi)^{3}}\left(\frac{1}{e^{\beta
      E_{\vec{p}}^{-}(\mu_{b})}+1} -\frac{1}{e^{\beta
      E_{\vec{p}}^{+}(\mu_{b})}+1}\right)\right]^{2}
\nonumber\\ &-&N_{f}\delta
(m+\eta)^{2}(n_{\pi}-1)G_{s}\left[\int\frac{d^{3}p}{(2\pi)^{3}E_{\vec{p}}}
  \left(1-\frac{1}{e^{\beta E_{\vec{p}}^{+}(\mu_{b})}+1}-\frac{1}{e^{\beta
      E_{\vec{p}}^{-}(\mu_{b})}+1}\right) \right]^{2}. 
\label{Veff2loopq3}
\end{eqnarray}
\end{widetext}

Adding Eqs.~(\ref{VeffMFOPT}), (\ref{Veff2loopq12}), and
(\ref{Veff2loopq3}), we get finally the  total effective potential in
the OPT expansion at order $\delta$, 

\begin{eqnarray}
 \mathcal{V}_{\delta^{1}}^{\mu,\mu_{b},\beta}(\sigma,\Delta,\eta,\alpha)
 &\equiv&V_{{\rm eff};1-{\rm
     loop},\delta^1}^{\mu,\mu_{b},\beta}(\sigma,\Delta,\eta,\alpha)
 \nonumber\\ &+& \tilde{V}_{{\rm eff};2-{\rm
     loop},\delta^1}^{\mu,\beta}(\eta,\alpha)
 \nonumber\\ &+&\bar{V}_{{\rm eff};2-{\rm
     loop},\delta^1}^{\mu_{b},\beta}(\eta). 
\label{VeffOPTtotal}
\end{eqnarray}

%%%%%%%%%%%%%%%%%%%%%%%%%%%%%%%%%%%%%%%%%%%%%%%%%%%%%%%%%%%%%%%%%%%%%%%%%
\subsection{The effective potential at zero temperature and 
finite chemical potential}

Since we are interested in describing the physics of dense and cold
matter,  we will from now on specialize on the expression for the
effective potential  at zero temperature. By taking the zero-temperature 
limit ($\beta\rightarrow \infty$) in Eq.~(\ref{VeffOPTtotal}),
we obtain, for each one of the terms in that equation, the  result

\begin{eqnarray}
V_{{\rm eff};1-{\rm loop},\delta^1}^{\mu,\mu_{b},\beta \rightarrow
  \infty}&=&\delta\frac{\sigma^{2}}{4G_{s}}
+\delta\frac{|\Delta|^{2}}{4G_{d}}-{\cal N}\left[J_{A}(\mu)
  \right.\nonumber\\ &+&\left.\delta(m+\eta)(\sigma-\eta)J_{E}(\mu)
  + \delta \alpha(\Delta-\alpha) J_{B}(\mu)\right] \nonumber\\ &-&
2N_{f}\left\{I_{A}+I_{C}(\mu_{b}) + \delta(m+\eta)(\sigma-\eta)
  \right.\nonumber\\ &\times&
  \left. \left[I_{D}-I_{E}(\mu_{b})\right]\right\}, \label{VeffMFOPT'T=0}
\end{eqnarray}

\begin{eqnarray}
\tilde{V}_{{\rm eff};2-{\rm loop},\delta^1}^{\mu,\beta \rightarrow
  \infty}&=&\frac{{\cal N} \delta}{4}
\left\{\left[(n_{\pi}+1)G_{s}-2G_{d}\right][J_{C}(\mu)]^{2}
\right. \nonumber\\ &-&
\left. (n_{\pi}+1)G_{s}\alpha^{2}[J_{B}(\mu)]^{2}\right\}
\nonumber\\ &-&\frac{{\cal N}
  \delta(m+\eta)^{2}}{4}\left\{\left[(n_{\pi}-1)G_{s}+2G_{d}\right]
            [J_{E}(\mu)]^{2}
            \right. \nonumber\\ &-&\left.(n_{\pi}-1)G_{s}
\alpha^{2}[J_{D}(\mu)]^{2}\right\},                     
\label{Veff2loopq12T=0}
\end{eqnarray}
and

\begin{eqnarray}
\bar{V}_{{\rm eff};2-{\rm loop},\delta^1}^{\mu_{b},\beta \rightarrow \infty}
&=&N_{f}\delta(n_{\pi}+1)G_{s}[I_{B}(\mu_{b})]^{2}-N_{f}\delta
(m+\eta)^{2} \nonumber\\ &\times&
(n_{\pi}-1)G_{s}[I_{D}-I_{E}(\mu_{b})]^{2},   \label{Veff2loopq3T=0}
\end{eqnarray}
where $I_{A}$ and $I_{D}$ correspond to the vacuum terms

\begin{eqnarray}
I_{A}&\equiv&\int \frac{d^{3}p}{(2\pi)^{3}}E_{\vec{p}}
\nonumber\\ 
&=&-\frac{1}{32\pi^{2}}\left\{M_{\eta}^{4}
\ln{\left[\frac{\left(\Lambda
        +\sqrt{\Lambda^{2}+M_{\eta}^{2}}
        \right)^{2}}{M_{\eta}^{2}}\right]}
  \right. 
\nonumber\\ 
&-&\left. 2\sqrt{\Lambda^{2}+
M_{\eta}^{2}}\, (2\Lambda^{3}+\Lambda
    M_{\eta}^{2})\frac{}{}\right\},
\end{eqnarray}

\begin{eqnarray}
I_{D}&\equiv&\int
\frac{d^{3}p}{(2\pi)^{3}}\frac{1}{E_{\vec{p}}}
\nonumber\\  &=&\frac{1}{4\pi^{2}}\left\{ \frac{}{}
  \Lambda\sqrt{\Lambda^{2}+M_{\eta}^{2}}
  \right. 
\nonumber\\ 
&-&\left.\frac{M_{\eta}^{2}}{2}\ln{\left[\frac{\left(\Lambda+\sqrt{\Lambda^{2}
          +M_{\eta}^{2}} \right)^{2}}{M_{\eta}^{2}}\right]}\right\},
\end{eqnarray}
where we have defined $M_\eta = m+\eta$ and we have explicitly
performed the integrals with a momentum cutoff $\Lambda$, whose value
will be fixed by fitting it together with the other parameters of the
model with the experimental observables (the pion mass, the pion decay
constant, and the quark condensate value).

The remaining terms, $I_{B}(\mu_{b})$, $I_{C}(\mu_{b})$, $I_{E}(\mu_{b})$, $J_{A}(\mu)$,
$J_{B}(\mu)$,  $J_{C}(\mu)$, $J_{D}(\mu)$,  and
$J_{E}(\mu)$ are the medium (chemical potential)--dependent terms,
given explicitly by the expressions

\begin{eqnarray}
I_{B}(\mu_{b})&\equiv&\int
\frac{d^{3}p}{(2\pi)^{3}}\Theta(\mu_{b}-E_{\vec{p}})
\nonumber\\  &=&\frac{\Theta(\mu_{b}-M_{\eta})}{6\pi^{2}}
\left(\mu_{b}^{2}-M_{\eta}^{2}\right)^{\frac{3}{2}},
\end{eqnarray}

\begin{eqnarray}
I_{C}(\mu_{b})&\equiv&\int \frac{d^{3}p}{(2\pi)^{3}}(\mu_{b}
-E_{\vec{p}})\Theta(\mu_{b}-E_{\vec{p}})
\nonumber\\ 
&=&\frac{\Theta(\mu_{b}-M_{\eta})}{32\pi^{2}}
\left\{M_{\eta}^{4}
  \ln{\left[\frac{\left(\sqrt{\mu_{b}^{2}-M_{\eta}^{2}}
        +\mu_{b}\right)^{2}}{M_{\eta}^{2}}\right]}
  \right. \nonumber\\ &+&\left. \frac{10}{3}\mu_{b}
\left(\mu_{b}^{2}-M_{\eta}^{2}\right)^{\frac{3}{2}}
  -2\mu_{b}^{3}\sqrt{\mu_{b}^{2}-M_{\eta}^{2}}
  \frac{}{}\right\},
\end{eqnarray}

\begin{eqnarray} 
I_{E}(\mu_{b})&\equiv&\int
\frac{d^{3}p}{(2\pi)^{3}}\frac{1}{E_{\vec{p}}}
\Theta(\mu_{b}-E_{\vec{p}})
\nonumber\\ &=&
\frac{\Theta(\mu_{b}-M_{\eta})}{4\pi^{2}}
\left\{\frac{}{}
  \mu_{b}\sqrt{\mu_{b}^{2}-M_{\eta}^{2}} \right. 
\nonumber\\ 
&-&  \left. \frac{M_{\eta}^{2}}{2}
\ln{\left[\frac{\left(\sqrt{\mu_{b}^{2}-M_{\eta}^{2}}
        +\mu_{b}\right)^{2}}{M_{\eta}^{2}}\right]}\right\},
\end{eqnarray}
and 

\begin{eqnarray}
J_{A}(\mu)&\equiv&\int
\frac{d^{3}p}{(2\pi)^{3}}[E_{\alpha\vec{p}}^{-}(\mu)
  +E_{\alpha\vec{p}}^{+}(\mu)], \\ 
J_{B}(\mu)&\equiv&\int
\frac{d^{3}p}{(2\pi)^{3}}\left[\frac{1}{E_{\alpha\vec{p}}^{-}(\mu)}
  +\frac{1}{E_{\alpha\vec{p}}^{+}(\mu)}\right],
\label{JB}
\\ J_{C}(\mu)&\equiv&\int \frac{d^{3}p}{(2\pi)^{3}}
\left[\frac{E_{\vec{p}}^{+}(\mu)}{E_{\alpha\vec{p}}^{+}(\mu)}
  -\frac{E_{\vec{p}}^{-}(\mu)}{E_{\alpha\vec{p}}^{-}(\mu)}\right],
\\  J_{D}(\mu)&\equiv&\int
\frac{d^{3}p}{(2\pi)^{3}}\frac{1}{E_{\vec{p}}}
\left[\frac{1}{E_{\alpha\vec{p}}^{+}(\mu)}-
\frac{1}{E_{\alpha\vec{p}}^{-}(\mu)}\right],
\\ J_{E}(\mu)&\equiv&\int
\frac{d^{3}p}{(2\pi)^{3}}\frac{1}{E_{\vec{p}}}
\left[\frac{E_{\vec{p}}^{-}(\mu)}{E_{\alpha\vec{p}}^{-}(\mu)}
  +\frac{E_{\vec{p}}^{+}(\mu)}{E_{\alpha\vec{p}}^{+}(\mu)}\right],
\end{eqnarray}
with the momentum integrations in the above expressions 
performed numerically, in practice (with the momentum cutoff $\Lambda$).

%%%%%%%%%%%%%%%%%%%%%%%%%%%%%%%%%%%%%%%%%%%%%%%%%%%%%%%%%%%%%%%%%%%%%%%%%
\section{Determination of parameters in the context of the OPT}
\label{sec5}

As already explained, the Lagrangian density in Eq.~(\ref{1LagNc3}) is an
effective model, and it is also nonrenormalizable, such that the
momentum cutoff $\Lambda$ used to regularize the momentum integrals,
which along with the quark current mass $m$ and the coupling constants
$G_s$ and $G_d$ (this last one will be treated as an independent
parameter, as mentioned earlier), must be chosen in such a way as to fit the
experimental data (most conveniently for vacuum quantities, i.e.,
when evaluated at zero temperature and chemical potential, $T = \mu =
0$). In the LN approximation, the procedure is very well understood
and explained in several places (see, e.g., Ref.~\cite{KLEVAN}). 
However, when using other nonperturbative methods, we are
led to possible corrections to these basic quantities, most notably
the pion mass and the pion decay constant, which are required to be
evaluated at the appropriate order according to the method used.  The
same is also true in the OPT method. How the fitting quantities change
in the context of the OPT was explained in detail in
Ref.~\cite{RUDOPT}. Here, for completeness, we will review and extend
the results of Ref.~\cite{RUDOPT} when the diquark interaction is also
present in the NJL Lagrangian density, as we have in
Eq.~(\ref{1LagNc3}).  This is an important step required for the
subsequent numerical analysis to be performed in the next
section and before one attempts to make predictions for other physical
quantities. We will start by first deriving consistently, in the OPT
method and at the order in which we are implementing our study, the basic
parameters from data.  These parameters, except for $G_{d}$ as already
mentioned,  can be estimated from the experimental data, i.e. the mass of
the pion $m_\pi$, the pion decay constant $f_\pi$, and the quark
condensate $\langle \bar{\psi}\psi \rangle$. 
{}For definiteness, the
values  for these quantities are set throughout this work to the
values $m_{\pi}=134 \ \mathrm{MeV}$,  $f_{\pi}=93 \ \mathrm{MeV}$, and
$-\langle \bar{\psi}\psi \rangle^{1/3}=250 \ \mathrm{MeV}$. 

Before discussing the appropriate fitting expressions, let us first
comment on the possible choices of values for the diquark coupling
constant $G_d$. Two possible constraints can be imposed in principle
on this constant: namely, that diquarks come to exist in the vacuum as
bound states and that they are stable,\footnote{In fact, 
the condition of diquark stability might not be a
necessary condition in principle, because it is not known whether 
the scalar diquark is really a bound state. (We thank L. He for 
pointing this out to us.)}  which implies that the
diquark mass $m_{d}$ must satisfy the condition
$0<m_{d}<2M_{q}$~\cite{SUNHE,ZHUANG,EBERTKLIM,EBERTKLIM2,EBERTKLIM3},
where $M_q$ is the effective quark mass.  
These two conditions can be translated in an lower
and upper limit
for $G_g$,
$G_{d}^{min}<G_{d}<G_{d}^{max}$,  where $G_{d}^{min}$ and
$G_{d}^{max}$ are determined by the
expressions~\cite{ZHUANG,EBERTKLIM}

\begin{eqnarray}
&&G_{d}^{max}=\frac{3}{2}G_{s}\frac{M_{q}}{M_{q}-m},
\\ 
&&G_{d}^{min}=\frac{\pi^{2}/4}{
\vphantom{\left(\frac{\Lambda+\sqrt{\Lambda^{2}
        +M_{q}^{2}}}{M_{q}}\right)}
  \Lambda\sqrt{\Lambda^{2}+M_{q}^{2}}+ M_{q}^{2} \, \ln \left(\frac{\Lambda
    +\sqrt{\Lambda^{2}+M_{q}^{2}}}{M_{q}}\right)}.
\end{eqnarray}
{}For the typical parameters provided by the LN approximation, from
the values for the mass of the pion $m_\pi$, the pion decay constant
$f_\pi$, and the quark condensate $\langle \bar{\psi}\psi \rangle$
given above, we find that $m \approx 4.99 \ \mathrm{MeV}$,
$M_{q}\approx 314 \ \mathrm{MeV}$,  $G_{s}\approx 4.94
\ \mathrm{GeV}^{-2}$, and $\Lambda\approx 653 \ \mathrm{MeV}$, which
give values for  $G_{d}$ in the range $0.81G_{s}\lesssim G_{d}
\lesssim 1.52G_{s}$.  However, in this work, 
we take these ranges of values for $G_d$ mostly as reference values.
Since we are mostly
interested in the study of the BEC-BCS crossover region, we find that at
values of $G_d$ around the minimum value $G_{d}^{min}$ there is no
BEC phase, the transition from the chiral phase to that of the 
condensate of diquarks is first order, preventing the appearance of
the BEC phase.  A large value of $G_d$ can make diquarks condense already
at very small values of the chemical potential. But diquark condensation
for chemical potential below the nucleon mass value is unrealistic,
so these cases should be excluded. This is in fact a  strong condition,
excluding the possibility of the BEC phase in the NJL, at least in its
simplest version.
We will say more about this when discussing our results in the next section.
In the present study, we find that a BEC phase
can appear in the LN case when $1.05G_{s}\lesssim G_{d} \lesssim 1.52G_{s}$.
More specifically, to allow comparison of our results with previous ones
obtained with the LN method and considered  in Ref.~\cite{SUNHE},
we will use values of $G_d$ such that $1.3G_{s}\lesssim G_{d} \lesssim 1.52G_{s}$,
which was also the same range of values considered in  Ref.~\cite{SUNHE}.
{}For the OPT
case, we also find allowed values of $G_d$ close to these in the case
of the absence of color neutrality. When the condition of color neutrality is
imposed, these values shift in the case of the OPT and give a much 
smaller window of values for $G_d$ allowing for a BEC phase, as we will show
in Sec.~\ref{OPTneutral}.

Let us now turn to the problem of determining the parameters
of the model.  The three basic parameters of the NJL model i.e., 
the values of the quark current mass $m$, the quark-antiquark coupling
$G_{s}$, and the ultraviolet cutoff $\Lambda$, are determined from the
system of equations, evaluated at the vacuum ($T=\mu=\mu_{b}=0$), formed
by Refs.~\cite{BUBA,KLEVAN}:
The system of equations is composed by the equation for the quark condensate

\begin{equation}
\sigma_c = -2G_s \langle {\bar \psi} \psi \rangle
\label{condzetapi2}
\end{equation}
by the pion mass equation, which is determined by the pole of the pion
propagator and by the equation for the pion decay
constant. Note that since all fitting expressions are
determined in the vacuum, where  the diquarks are not condensed i.e.,
$\Delta_{c}=0$ the presence of a diquark interaction will not affect
the fitting parameters, at least in the LN approximation, where
diquark fluctuations do not contribute. This is, however, not true
in the OPT case, where already at order $\delta$ there will be
two-loop terms with diquark fluctuations contributing to both the 
pion mass and the pion decay constant. Thus, the fittings
in the OPT case will depend explicitly on the diquark coupling
$G_d$, as we will show below. 
The other parameters can be found by solving a  system
of equations formed by the gap equation determining
the chiral condensate $\sigma_c$, 

\begin{equation}
\frac{\partial V_{\rm eff}}{\partial \sigma} \Bigr|_{\sigma=\sigma_c}=0,
\label{gapsigma}
\end{equation}
and the diquark condensate,

\begin{equation}
\frac{\partial V_{\rm eff}}{\partial \Delta} \Bigr|_{\Delta=\Delta_c}=0,
\label{gapdelta}
\end{equation}
and, in the OPT case, by the two PMS equations (\ref{pmseta}) and
(\ref{pmsalpha}) used to determine the optima $\bar \eta$ and $\bar \alpha$.
Note also that, in the OPT case, in
the vacuum, since the value of  $\Delta$ that minimizes the
effective  potential is $\Delta_{c}=0$, it can be easily shown
that the PMS Eq.~(\ref{pmsalpha})  for $\alpha$ provides a value
$\bar{\alpha}=0$. In practice, this means that we can get  all the
vacuum equations for the parameter calculations from the thermodynamic
effective potential,

\begin{equation}
V_{\rm eff}^{(vac)}(\sigma_{c},\bar{\eta})
=\mathcal{V}_{\delta^{1}}^{\mu=0,\mu_{b}=0,\beta\rightarrow\infty}
(\sigma_{c},\Delta_{c}=0,\bar{\eta}),
\end{equation} 
which is found after we make
the substitutions in, e.g., Eq.~(\ref{VeffOPTtotal}):
$[\alpha,E_{\alpha \vec{p}}^{\pm}(\mu)] \rightarrow [0,E_{\vec{p}}^{
    \pm}(\mu)]$,  $\mu=\mu_{b}=0$, $\Delta=0$, and $\beta\rightarrow \infty$, which gives the OPT
expression for the effective potential, at order $\delta$ and in the
vacuum,

\begin{eqnarray} \label{Omegavac}
V_{\rm eff}^{(vac)}(\sigma_{c},\bar{\eta})&=&\delta\frac{\sigma_{c}^{2}}{4G_{s}}
 -2N_{c}N_{f}[I_{A}+\delta (m+\bar{\eta})(\sigma_{c}-\bar{\eta})I_{D}]
 \nonumber \\ &-&N_{c}N_{f}\delta (m+\bar{\eta})^{2}G_{s}
 \nonumber\\ &\times&\left[(n_{\pi}-1)+2\frac{(N_{c}-1)}{N_{c}}\frac{G_{d}}{G_{s}}\right]I_{D}^{2}.
\end{eqnarray}

The gap equation (\ref{gapsigma})  for $\sigma_{c}$, the relation with
the chiral condensate and the PMS Eq.~(\ref{defPMS})  to $\bar{\eta}$
are easily obtained from Eq.~(\ref{Omegavac}) and they result in

\begin{eqnarray}
&&M_{q}^{OPT} = m+4G_{s}N_{c}N_{f}\mathcal{M}I_{2}, 
\label{SysgapOPT}\\
&&\langle\bar{\psi}\psi\rangle  =  -\frac{M_{q}^{OPT}-m}{4G_{s}}, 
\label{SyscondOPT} \\
&&\mathcal{M} = M_{q}^{OPT}+f(G_{d})G_{s}\mathcal{M}I_{2}, 
\label{SysPMSOPT}
\end{eqnarray}
where we have defined $M_{q}^{OPT}=m+\sigma_{c}$,
$\mathcal{M}=m+\bar{\eta}$,

\begin{eqnarray} \label{deff(Gd)}
f(G_{d})&\equiv&(n_{\pi}-1)+\frac{2G_{d}}{G_{s}}\frac{(N_{c}-1)}{N_{c}},
\end{eqnarray}
and $I_2$ in Eq.~(\ref{SysPMSOPT}) is given by

\begin{eqnarray}
I_{2}&\equiv&\int \frac{d^{3}p}{(2\pi)^{3}}\frac{1}{E_{\vec{p}}}
\nonumber \\ 
&=&\frac{1}{4\pi^{2}}\left\{\frac{}{}
  \Lambda \sqrt{\Lambda^{2}+\mathcal{M}^{2}}-\frac{\mathcal{M}^{2}}{2}
  \right. 
\nonumber\\ 
&\times&  \left. \ln\left[\frac{(\Lambda+\sqrt{\Lambda^{2}
        +\mathcal{M}^{2}})^{2}}{\mathcal{M}^{2}}\right]\right\} 
\label{Int_2}.
\end{eqnarray}

The equations for the pion  mass  $m_{\pi}$ and for the pion decay
constant $f_{\pi}$ are evaluated next in the context of the OPT
approximation.

Note that from Eqs.~(\ref{SysgapOPT}) and (\ref{SysPMSOPT}), we obtain the
simple relation between $\bar \eta$ and $\sigma_c$ in the vacuum:

\begin{eqnarray}
{\bar \eta} &=& \sigma_c [1+f(G_d)/(4N_c N_f)]
\nonumber \\
&=& \sigma_c \left[ 1+ \frac{n_\pi-1}{4 N_c N_f} 
+\frac{G_d}{G_s} \frac{(N_c-1)}{2 N_c^2 N_f} \right],
\end{eqnarray}
which shows that in the large-$N_c$ limit we reproduce the result
$\bar \eta = \sigma_c$ as expected in the LN approximation.

%%%%%%%%%%%%%%%%%%%%%%%%%%%%%%%%%%%%%%%%%%%%%%%%%%%%%%%%%%%%%%%%%%%%%%%%%%
\subsection{The pion mass equation}
\label{pionmass:Sec}

The  pion mass is determined by the pole of the  pion propagator,
which can be expressed as~\cite{RUDOPT} 

\begin{eqnarray} \label{InversPropPionOPT}
1-2G_{s}\Pi_{\pi}(q^{2}),
\end{eqnarray}
where $\Pi_{\pi}(q^{2})$ is the pion self-energy, evaluated
consistently  at the required OPT order. In our case, where we are
evaluating quantities up to ${\cal O}(\delta)$ in the OPT expansion,
we will have contributions to the pion self-energy that include both
one- and two-loop terms, which are  shown in {}Fig.~\ref{fig:
  FDparameters}.

%%%%%%%%%%%%%%%%%%%%%%%%%%%%%%%%%%%%%%%%%%%%%%%%%%%%%%%
\begin{figure*}[t]
\includegraphics[width=7cm]{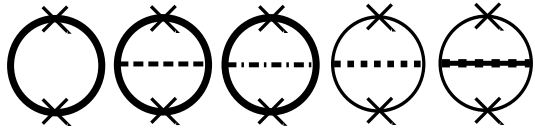}
\caption[dd]{{\footnotesize Diagrams relevant to the calculation of
    the  pion mass and its decay constant in the OPT expansion up to
    ${\cal O}(\delta)$.  The thick continuous  line represents vacuum
    fermionic propagators for quarks with all colors, which are
    evaluated when  $\Delta=\Delta_{c}=0$; the thin line represents
    vacuum fermionic propagators for only   quarks with colors 1 and
    2; the dashed line represents the chiral bosonic scalar $\zeta$
    field propagator;  the dash-dotted one represents the pion
    $\vec{\pi}$-fields propagator; the dotted one is related  to the
    $\phi_{R}$ field; and the continuous-dotted line is associated with
    the $\phi_{I}$ field.  In addition, the vertex pairs represented
    by an ``X'' in each diagram can be  $\gamma^{5}\tau_{i,j}$ (in the
    case of the pion mass equation) or
    $\gamma^{5}\gamma_{\mu,\nu}\tau_{i,j}$ (in the case of the decay
    constant equation).  All quantities are calculated when
    $\delta=1$.}}
\label{fig: FDparameters}
\end{figure*}
%%%%%%%%%%%%%%%%%%%%%%%%%%%%%%%%%%%%%%%%%%%%%%%%%%%%%%%%%%%%%

The free fermion propagators shown in {}Fig.~\ref{fig: FDparameters}
and related to the quarks $q_1$ and $q_2$ (red and green in color space)
and to $q_{3}$ are given, respectively, by

\begin{eqnarray}
iG_{\Psi}^{vac}(p)&=& i  G^{vac}(p)1_{NG}, \label{GpOPTPsivac}
\end{eqnarray}
and

\begin{eqnarray}
iG_{q_{3}}^{vac}(p)&=& i G^{vac}(p), \label{GpOPTq3vac}
\end{eqnarray}
where

\begin{eqnarray}
G^{vac}(p)=\frac{\slashed{p}+\mathcal{M}}{p^2-\mathcal{M}^2+i\epsilon}, 
\end{eqnarray}
and $\Psi$ represents the quarks $q_1$ and $q_2$ in the Nambu-Gor'kov
space~\cite{gorkov,Nambu:1960tm}, with $1_{NG}$ being the identity matrix in this space.

The one-loop diagram shown in {}Fig.~\ref{fig: FDparameters}, when
using the vertex $\tau_{i}\gamma^{5}$ and the {}Feynman rules
obtained from  Eq.~(\ref{LBOPTestNJL}), can be written
explicitly in the form

\begin{eqnarray}
i\Pi_{\pi}^{(1)}(q^{2})\delta_{ij}&=&-\int\frac{d^4p}{(2\pi)^{4}}
\mathrm{Tr}_{c,f,D}\left[iG^{vac}(p)(\tau_{i}\gamma^{5})
  \right. \nonumber\\ &\times&
  \left. iG^{vac}(p+q)(\tau_{j}\gamma^{5})\right],
\end{eqnarray}
where $q$ denotes here the external momentum and $\delta=1$ is considered in this and in
all subsequent terms evaluated in the OPT expansion.  After we
perform the traces in flavor and color spaces, we find

\begin{eqnarray}
\Pi_{\pi}^{(1)}(q^2)&=&2iN_{c}N_{f}\left[2I_{G}(\mathcal{M})-q^{2}I(q^2)\right], 
\label{Pi1loop}
\end{eqnarray}
where

\begin{eqnarray}
I_{G}(\mathcal{M})=\int\frac{d^{4}p}{(2\pi)^{4}}
\frac{1}{p^{2}-\mathcal{M}^{2}+i \epsilon}
  =-\frac{i}{2}I_{2} ,
\label{IG}
\end{eqnarray}
and 

\begin{eqnarray}
\!\!\!\!\!\!\!\!\!\!\!\!I(q^{2})\!=\!\!
\int\!\frac{d^{4}p}{(2\pi)^{4}}
\frac{1}{(p^{2}\!-\!\mathcal{M}^{2}\!+\!i\epsilon)
[(p\!+\!q)^{2}\!-\!\mathcal{M}^{2}+i\epsilon]}. 
\label{I(q2)}
\end{eqnarray}

The self-energy terms generated by the two-loop diagrams, given by the
second and third diagrams shown in {}Fig.~\ref{fig: FDparameters} and
related to  the scalar  $\zeta$ and pion $\vec{\pi}$  chiral fields
can be written, respectively, as

\begin{eqnarray}\label{Pisigmabruto}
-i\Pi_{\pi \ ij}^{(2),\zeta}(q^{2})&=&-i2G_{s}\int_{p_{1},p_{2}}
\mathrm{Tr}_{c,f,D} \left[(-i)iG^{vac}(p_{1})(\tau_{i}\gamma^{5})
  \right. \nonumber\\ &\times& \left. iG^{vac}(p_{1}+q)
  (-i)iG^{vac}(p_{2}+q) \right. \nonumber\\ &\times&
  \left. (\tau_{j}\gamma^{5})iG^{vac}(p_{2})\right], \end{eqnarray}
and

\begin{eqnarray}\label{Pipionbruto}
-i\Pi_{\pi \ ij}^{(2),\vec{\pi}}(q^{2})&=&-i2G_{s}\int_{p_{1},p_{2}}
\mathrm{Tr}_{c,f,D} \left[(\tau_{k}\gamma^{5})iG^{vac}(p_{1})
  \right. \nonumber\\ &\times&
  \left. (\tau_{i}\gamma^{5})iG^{vac}(p_{1}+q)(\tau_{k}\gamma^{5})iG^{vac}(p_{2}+q)
  \right. \nonumber\\ &\times&\left. (\tau_{j}\gamma^{5})iG^{vac}(p_{2})\right],
\end{eqnarray}
where

\begin{eqnarray}
\int_{p_{1},p_{2}} \equiv
\int\frac{d^4p_{1}}{(2\pi)^{4}}\frac{d^4p_{2}}{(2\pi)^{4}}.
\end{eqnarray}

Evaluating again the traces in the above expressions, we obtain

\begin{eqnarray}
\Pi_{\pi}^{(2),\zeta}(q^{2})&=&-8G_{s}N_{c}N_{f}
\left\{\frac{}{}I_{G}^{2}(\mathcal{M})-q^{2}\left[
  \vphantom{\frac{q^2}{4}}I_{G}(\mathcal{M})I(q^2)
  \right. \right. \nonumber\\ &-&
  \left. \left. \mathcal{M}^{2}I^{2}(q^{2})-\frac{q^2}{4}I^{2}(q^{2})\right]
\right\},
\label{Pisigma}
\end{eqnarray}
and 

\begin{eqnarray}
\Pi_{\pi}^{(2),\vec{\pi}}(q^{2})&=&-8(n_{\pi}-2)G_{s}N_{c}N_{f}
\left\{\frac{}{}I_{G}^{2}(\mathcal{M})  \right. \nonumber\\  &-&
q^{2}\left. \left[I_{G}(\mathcal{M})I(q^2)+\mathcal{M}^{2}I^{2}(q^{2})
  -\frac{q^2}{4}I^{2}(q^{2})\right]\right\} \ ,
\nonumber\\ \label{Pipion}
\end{eqnarray}
where

\begin{eqnarray}
\Pi_{\pi \ ij}^{(2),\zeta}(q^{2})&=&\Pi_{\pi}^{(2),\zeta}(q^{2})
\delta_{ij}, \nonumber\\ \Pi_{\pi \ ij}^{(2),\vec{\pi}}(q^{2})&=&
\Pi_{\pi}^{(2),\vec{\pi}}(q^{2}) \delta_{ij}.
\end{eqnarray}

The contributions of the last two diagrams shown in {}Fig.~\ref{fig:
  FDparameters} are related to the real and imaginary components
of the diquark scalar field,  $\phi_{R}$ and $\phi_{I}$, respectively.
In this case, only the vacuum propagator relative to the
Nambu-Gor'kov spinor $\Psi$, given by Eq.~(\ref{GpOPTPsivac}), needs to be taken
into account.  Explicitly, we have that

\begin{widetext}
\begin{eqnarray}\label{PiphiRbruto}
-i\Pi_{\pi
  \ ij}^{(2),\phi_{R}}(q^{2})&=&\frac{1}{2}(-i2G_{d})\int_{p_{1},p_{2}}
\mathrm{Tr}_{all}\left[i^{2}\gamma^{5}\tau_{2}t_{2}\begin{pmatrix} 0&1
    \\ 1&0
\end{pmatrix} iG_{\Psi}^{vac}(p_{1})\gamma^{5}\begin{pmatrix} \tau_{i}&0 \\ 0&\tau_{i}^{T}
\end{pmatrix} iG_{\Psi}^{vac}(p_{1}+q)\right. \nonumber\\
&\times& \left.  i^{2}\gamma^{5}\tau_{2}t_{2} \begin{pmatrix} 0&1
    \\ 1&0
\end{pmatrix}   iG_{\Psi}^{vac}(p_{2}+q) \gamma^{5}\begin{pmatrix} \tau_{j}&0 \\ 0&\tau_{j}^{T}
\end{pmatrix} iG_{\Psi}^{vac}(p_{2})\right],
\end{eqnarray}
and 

\begin{eqnarray}\label{PiphiIbruto}
-i\Pi_{\pi
  \ ij}^{(2),\phi_{I}}(q^{2})&=&\frac{1}{2}(-i2G_{d})\int_{p_{1},p_{2}}
\mathrm{Tr}_{all}\left[(-i^{2}\gamma^{5}\tau_{2}t_{2}) \begin{pmatrix}
    0&-i \\  i&0\end{pmatrix}
    iG_{\Psi}^{vac}(p_{1})\gamma^{5}\begin{pmatrix} \tau_{i}&0
      \\ 0&\tau_{i}^{T}
\end{pmatrix}iG_{\Psi}^{vac}(p_{1}+q)\right. \nonumber\\
&\times& \left.  (-i^{2}\gamma^{5}\tau_{2}t_{2}) \begin{pmatrix} 0&-i
      \\ i&0
\end{pmatrix}   iG_{\Psi}^{vac}(p_{2}+q) \gamma^{5}\begin{pmatrix} \tau_{j}&0 \\ 0&\tau_{j}^{T}
\end{pmatrix} iG_{\Psi}^{vac}(p_{2})\right],
\end{eqnarray}
\end{widetext}
where $\mathrm{Tr}_{all}$ also incorporates the trace in the
Nambu-Gor'kov space.

It is easy to show that  $\Pi_{\pi
  \ ij}^{(2),\phi_{R}}(q^{2})=\Pi_{\pi \ ij}^{(2),\phi_{I}}(q^{2})$.
Therefore, we only have to calculate Eq.~(\ref{PiphiRbruto}) or, equivalently,
Eq.~(\ref{PiphiIbruto}).  After the calculation of the traces, we obtain
the joint contribution of the two diagrams:
  
\begin{eqnarray}
\Pi_{\pi}^{(2),\phi}(q^{2})&\equiv&\Pi_{\pi}^{(2),\phi_{R}}(q^{2})
+\Pi_{\pi}^{(2),\phi_{I}}(q^{2})  \nonumber \\ &=&
2\Pi_{\pi}^{(2),\phi_{R}}(q^{2})\nonumber\\ &=&-16 G_{d} (N_{c}-1)N_{f}\left\{\frac{}{}I_{G}^{2}(\mathcal{M})
\right. \nonumber\\ &-&
\left. q^{2}\left[I_{G}(\mathcal{M})I(q^2)+\mathcal{M}^{2}I^{2}(q^{2})
  -\frac{q^2}{4}I^{2}(q^{2})\right]\right\}. 
\nonumber\\ \label{Pidiquark}
\end{eqnarray}

The pion mass $m_{\pi}$ is the pole of its propagator. This means that
Eq.~(\ref{InversPropPionOPT}) should be null when we make
$q^{2}\rightarrow m_{\pi}^{2}$.  Since $\Pi_{\pi}(q^{2})$ is the
sum of Eqs.~(\ref{Pi1loop}), (\ref{Pipion}), (\ref{Pisigma}), and
(\ref{Pidiquark}), we can write

\begin{eqnarray}
0&=&1-2G_{s}\Pi_{\pi}(m_{\pi}^{2})
\nonumber\\ 
&=&1-4iG_{s}N_{c}N_{f}\left[2I_{G}(\mathcal{M})-m_{\pi}^{2}I(m_{\pi}^2)\right]
\nonumber \\
&+& 4 N_{c}N_{f}G^{2}_{s} 
 \left\{f(G_{d})\left[2I_{G}(\mathcal{M})
  -m_{\pi}^{2}I(m_{\pi}^{2})\right]^2
\right.
\nonumber \\
&-& \left.
  4\left[f(G_{d})-2\right]m_{\pi}^{2}\mathcal{M}^{2}I^{2}(m_{\pi}^{2})
\right\}\;, 
\label{IRPP}
\end{eqnarray}
where $f(G_{d})$ was already defined in Eq.~(\ref{deff(Gd)}), and the integral
$I(m_{\pi}^2)$  is given by
           
\begin{eqnarray}
I(m_{\pi}^{2})&=&\frac{i}{8\pi^{2}}\left[\mathrm{ln}\left(\frac{\Lambda
    +\sqrt{\Lambda^{2}+\mathcal{M}^{2}}}{\mathcal{M}}\right)
  -\sqrt{4\frac{\mathcal{M}^{2}}{m_{\pi}^{2}}-1}
  \right. \nonumber\\ &\times&\left. \mathrm{tan}^{-1}\left(\frac{\Lambda}{\sqrt{\Lambda^{2}
      +\mathcal{M}^{2}}\sqrt{4\mathcal{M}^{2}/m_{\pi}^{2}-1}}\right)\right]. 
\nonumber \\ 
\label{Int_4}
\end{eqnarray}

Now, iterating once the PMS equation (\ref{SysPMSOPT}) and
substituting in  the gap equation (\ref{SysgapOPT}), we get the
relation

\begin{eqnarray}
\frac{m}{M_{q}^{OPT}}&=&1-8iG_{s}N_{c}N_{f}I_{G}(\mathcal{M})
\nonumber\\ &+&16f(G_{d})G_{s}^{2}N_{c}N_{f}I_{G}^{2}(\mathcal{M}), \label{gapPMS}
\end{eqnarray}
which, when inserted into Eq.~(\ref{IRPP}), gives us the result

\begin{eqnarray}
\frac{m}{M_{q}^{OPT}}&=&4G_{s}N_{c}N_{f}m_{\pi}^{2}\left\{
\vphantom{\left(\mathcal{M}^{2}-\frac{m_{\pi}^2}{4}\right)}
-iI(m_{\pi}^{2})+4f(G_{d})G_{s}
\right. \nonumber\\ &\times&\left. \left[I_{G}(\mathcal{M})I(m_{\pi}^{2})
  +\left(\mathcal{M}^{2}-\frac{m_{\pi}^2}{4}\right)I^{2}(m_{\pi}^{2})\right]
\right. \nonumber\\ &-&
\left. 8G_{s}\mathcal{M}^{2}I^{2}(m_{\pi}^{2})\vphantom{\left(\mathcal{M}^{2}
  -\frac{m_{\pi}^2}{4}\right)}\right\}. 
\label{SysmpionOPT}
\end{eqnarray}
We can clearly see that Eq.~(\ref{SysmpionOPT}) satisfies the
Goldstone theorem.  When we take $m = 0$ (the chiral case) in
Eq.~(\ref{SysmpionOPT}), we automatically obtain $m_{\pi}=0$,
consistent with the Goldstone theorem.

%%%%%%%%%%%%%%%%%%%%%%%%%%%%%%%%%%%%%%%%%%%%%%%%%%%%%%%%%%%%%%%%%%%%%%%%%
\subsection{The pion decay constant equation}

Let us now evaluate the pion decay constant in the OPT expansion to
order $\delta$. The  pion decay constant can be expressed
as~\cite{RUDOPT}

\begin{eqnarray}
\langle 0 | T
A_{\mu}^{i}(q)A_{\nu}^{j}(0)|0\rangle=ig_{\mu\nu}\delta^{ij}f_{\pi}^{2}
+\mathcal{O}(q_{\mu}q_{\nu}),
\end{eqnarray}
where $A_{\mu}^{i}\equiv
\bar{\psi}\gamma_{\mu}\gamma_{5}(\tau^{i}/2)\psi$.  In practice, we
can take advantage of all the diagrams of {}Fig.~\ref{fig:
  FDparameters} again,  but we replace the vertex $\gamma^{5}\tau_{i,j}$ with
$\gamma^{5}\gamma_{\mu}\tau_{i,j}/2$  to compute $f_{\pi}^{2}$. Since
the calculations are analogous, yet more laborious than  those made
previously to obtain the expression containing the pion mass, we show some the
details in the Appendix. {}From the
results given there,  we extract that the  contribution from each loop term
contributing to $f_{\pi}$ can be expressed in the form

\begin{eqnarray}
f_{\pi,1}^{2}&=&-2iN_{c}N_{f}\mathcal{M}^{2}I(0),
 \label{fpi1loopfinal}
\end{eqnarray}

\begin{eqnarray}
f_{\pi,\zeta}^{2}&=&4G_{s}N_{c}N_{f}\mathcal{M}^{4}I^{2}(0),
 \label{fpisfinal}
\end{eqnarray}

\begin{eqnarray}
f_{\pi,\vec{\pi}}^{2}&=&4G_{s}N_{c}N_{f}(n_{\pi}-2)\mathcal{M}^{4}I^{2}(0),
 \label{fpipsfinal}
\end{eqnarray}

\begin{eqnarray}
f_{\pi,\phi}^{2}&=&8G_{d}(N_{c}-1)N_{f}\mathcal{M}^{4}I^{2}(0), 
\label{fpidfinal}
\end{eqnarray}
with the integral $I(0)$ obtained from the limit  $q^{2}\rightarrow 0$
applied to Eq.~(\ref{I(q2)}), which gives
 
\begin{eqnarray}
I(0)&=&\frac{i}{8\pi^{2}}\left[\mathrm{sinh}^{-1}\left(\frac{\Lambda}{\mathcal{M}}\right)
  -\frac{\Lambda}{\sqrt{\Lambda^{2}+\mathcal{M}^{2}}}\right].
\end{eqnarray}

The final expression for $f_{\pi}^{2}$ is obtained by adding
Eqs.~(\ref{fpi1loopfinal}),  (\ref{fpisfinal}), (\ref{fpipsfinal}), and
(\ref{fpidfinal}), which finally gives

\begin{eqnarray} \label{SysfpiOPT}
f_{\pi}^{2}&=&-2iN_{c}N_{f}\mathcal{M}^{2}I(0)+4N_{c}N_{f}G_{s}f(G_{d})
\mathcal{M}^{4}I^{2}(0). \nonumber\\
\end{eqnarray}

%%%%%%%%%%%%%%%%%%%%%%%%%%%%%%%%%%%%%%%%%%%%%%%%%%%%%%%%%%%%%%%%
\subsection{The complete fitting expressions in the OPT expansion to ${\cal O}(\delta)$}

The complete set of consistent equations that need to be solved in order to
provide the values of the parameters, once the numerical data for
$m_\pi$, $f_\pi$, and $\langle \bar q q\rangle$ are provided, is then

\begin{eqnarray}
\label{MqOPT}
&&M_{q}^{OPT}=m+4G_{s}N_{c}N_{f}\mathcal{M}I_{2},
  \\ 
\label{calM}
&&\mathcal{M}=M_{q}^{OPT}+f(G_{d})G_{s}\mathcal{M}I_{2},
  \\ &&\langle\bar{\psi}\psi\rangle = -\frac{M_{q}^{OPT}-m}{4G_{s}},
  \\ &&f_{\pi}^{2}=-2iN_{c}N_{f}\mathcal{M}^{2}I(0)+4f(G_{d})N_{c}N_{f}G_{s}
  \mathcal{M}^{4}I^{2}(0), \nonumber\\
\end{eqnarray}
and

\begin{eqnarray}
\frac{m}{M_{q}^{OPT}}&=&4G_{s}N_{c}N_{f}m_{\pi}^{2}\left\{\vphantom{
  \left(\mathcal{M}^{2}-\frac{m_{\pi}^2}{4}\right)}
-iI(m_{\pi}^{2})+4f(G_{d})G_{s}
\right. \nonumber\\ &\times&\left. \left[I_{G}(\mathcal{M})I(m_{\pi}^{2})
  +\left(\mathcal{M}^{2}-\frac{m_{\pi}^2}{4}\right)I^{2}(m_{\pi}^{2})\right]
\right. \nonumber\\ &-&
\left. 8G_{s}\mathcal{M}^{2}I^{2}(m_{\pi}^{2})\vphantom{\left(\mathcal{M}^{2}
  -\frac{m_{\pi}^2}{4}\right)}\right\} \ .
\end{eqnarray}

%%%%%%%%%%%%%%%%%%%%%%%%%%%%%%%%%%%%%%%%%%%%%%%%%%%%%%%
\begin{table}[!htpb]
\caption{\footnotesize Parameter values used in the OPT scheme
and in the LN approximation (last line). The inputs used are $m_{\pi}=134 
\ \mathrm{MeV}$, $f_{\pi}=93
\ \mathrm{MeV}$,  and $-\langle \bar{\psi}\psi \rangle^{1/3}=250
\ \mathrm{MeV}$.}
\label{TAB:PARAMETROS}
\begin{tabular}{ccccc}
\hline $G_{d}/G_{s}$ & $ M_{q} \ (\mathrm{MeV})$ & $m
\ (\mathrm{MeV})$  & $G_{s} \ (\mathrm{GeV}^{-2})$ & $\Lambda
\ (\mathrm{MeV})$\\ \hline 
$1.3$  & $293.445$  & $4.777$ & $4.619$ & $640.112$
\\ 
$1.4$  & $292.973$  & $4.760$ & $4.611$ & $639.597$
\\ 
$1.5$  & $292.520$  & $4.744$ & $4.604$ & $639.077$
\\  
1.53  & 292.389 & 4.739  & 4.602  & 638.920
\\ 1.54     & 292.345    & 4.737 & 4.602        & 638.867
\\ 1.55     & 292.302    & 4.735 & 4.601        & 638.815
\\  \hline
LN   & $313.519$  & $4.987$ & $4.937$ & $653.331$
\\ \hline
\end{tabular}
\end{table}
%%%%%%%%%%%%%%%%%%%%%%%%%%%%%%%%%%%%%%%%%%%%%%%%%%%%%%%%%

{}From the input values, we obtain numerically, sets of parameters for some
values of $G_{d}/G_{s}$, as shown in Table~\ref{TAB:PARAMETROS}. 
Note that, as compared to the  LN
approximation, the corrections due to OPT cause a slight
drop\footnote{Translating in  percentages, there is a decrease of
  approximately $2\%$ in $\Lambda$, $6\%$  to $7\%$ in $G_{s}$, $3\%$ to
  $5\%$ in $m$, and $6\%$ to $7\%$ in $M_{q}$.} in  all parameters of
the table, and this fall is intensified with increasing coupling
between quarks (represented by $G_{d}$). The  LN approximation, which
can be  obtained when we neglect the OPT two-loop contributions in the
equations that  compose the system, does not provide parameters that
depend on $G_{d}$, as already explained, since diquark fluctuations would
contribute with subleading $1/N_c$ correction terms, but these terms
do contribute in the OPT case.  The values corresponding to the ratios
$G_s/G_d=1.53,\, 1.54$, and $1.55$ and shown in the
Tab.~\ref{TAB:PARAMETROS} will be used when imposing the color
neutrality condition, while the other values will be used in the
absence of color neutrality and in the comparison of our OPT results
with those obtained in the LN approximation.

%%%%%%%%%%%%%%%%%%%%%%%%%%%%%%%%%%%%%%%%%%%%%%%%%%%%%%%%%%%%%%%%%%%%%%%%
\section{Numerical results using the OPT for the cold and dense system}
\label{results}

Before presenting our results, it is useful to first recall a few
properties regarding the BEC-BCS crossover and the requirement
for color neutrality.

%%%%%%%%%%%%%%%%%%%%%%%%%%%%%%%%%%%%%%%%%%%%%%%%%%%%%%%%%%%%%%%%%%%%%%
\subsection{The BEC-BCS crossover}

If we start from the dispersion relation e.g., the one from the mean
field LN approximation, from Eq.~(\ref{fermion-E}) and set
$\delta=0$, $\mu'=\mu$ (red and green quarks), $\alpha_s\to \Delta$, and $\eta \to \sigma$, 
we have, for example, $E_{\Delta,\vec{p}}^{-}(\mu)=\sqrt{\left[\sqrt{p^2+M_q^2}-\mu\right]^{2}
  +\Delta^2}$. {}For small chemical potential $\mu\leq M_q$, the
minimum of the dispersion is located at $|\vec{p}| = 0$, with particle
gap energy $\sqrt{M_q^2 +  \Delta^2}$, which would correspond to the
fermionic (quark) spectrum in the BEC state. At  values of chemical
potential such that $\mu>M_q$, the minimum of the dispersion is
shifted to  $|\vec{p}| \neq 0 $, and the particle gap is $\Delta$. This
corresponds to the fermionic spectrum in the BCS state.  It is then
useful to define an effective chemical potential
$\mu_{N}\equiv\mu-M_{q}$, which will serve as an indicator of the
BEC-BCS crossover~\cite{SUNHE}.

%%%%%%%%%%%%%%%%%%%%%%%%%%%%%%%%%%%%%%%%%%%%%%%%%%%%%%%%%%%%%%%%%%%%
\subsection{Color neutrality condition}

In the model given by Eq.~(\ref{1LagNc3}) for $N_{c}=3$, in the choice that allows
only red and green  color quarks to form diquarks and that leaves out
the blue ones, for example,  it follows that when equal chemical
potentials are introduced for the three colors
$\mu_{r}=\mu_{g}=\mu_{b}\equiv \mu_{B}/3$, where $\mu_{B}$ represents the baryon 
chemical potential the phase characterized by the absence of the
diquark condensate, $\Delta_c=0$,  keeps the color symmetry  $SU(3)$,
while the phase at which the condensate is nonzero, $\Delta_c\neq 0$,
 breaks  the $SU(3)$ color symmetry down to $SU(2)$.  However, in
the latter case, the number densities of the quarks that form the
diquarks,  $n_{r}$ and $n_{g}$, are identical and are larger than the
density of the blue-colored  quarks,
$n_{b}$~\cite{BubaShov,Diet,EBERTKLIM2,SUNHE}. This means that in this
phase, the  system as a whole does not have the property of color
neutrality, which is physically verified.  In fact, such a situation
also occurs in QCD when we consider the two-flavor superconducting
color phase,  but it is possible to generate the eighth gluon field,
which guarantees the color  neutrality automatically~\cite{EBERTKLIM2}
in theory. Effectively, it generates a chemical  potential
$\mu_{8}$. Since in the NJL model we do not have gluon degrees of
freedom,  what is done to ensure color neutrality is to add by hand a
chemical potential term $\mu_{8}$ in  the Lagrangian density of the
theory -- $\mu_{8}\bar{\psi}\gamma^{0}T_{8}\psi$, with
$T_{8}=\sqrt{3}\lambda_{8}$ -- and impose that $\langle Q_{8}\rangle=0$,
which is equivalent  to demanding the condition

\begin{eqnarray} \label{neutralitycondition}
n_{8}=-\frac{\partial \Omega}{\partial \mu_{8}}=0,
\end{eqnarray}
where $\Omega$  is the thermodynamic potential in the desired
approximation.  {}Furthermore, in practice, the chemical potential
$\mu_{8}$ enters in the final  expressions obtained. Until then, we can
simply make the changes~\cite{SUNHE}  $\mu_{g}=\mu_{r}=\mu \rightarrow
\mu_{B}/3+\mu_{8}/3$ and $\mu_{b} \rightarrow \mu_{B}/3-2\mu_{8}/3$.  This will be
the procedure we will also follow here when demanding color
neutrality.

Note that besides the imposition of color neutrality,  
electric charge neutrality in principle should also be considered.
Including electric charge neutrality introduces
an extra chemical potential, $\mu_Q$, which is proportional to the electric charges
for the $u$ and $d$ quarks and introduces an explicit difference in
chemical potentials for these quarks. This difference in chemical potentials
can lead to some important effects, such as a gapless color 
superconducting phase~\cite{Huang:2002zd,Shovkovy:2003uu}.
Since in this work we are primarily interested in the comparison of the
LN results for the BEC-BCS crossover with the OPT ones, 
we will here neglect for simplicity
the condition of electric charge neutrality, as was the case also in 
the previous works~\cite{EBERTKLIM2,SUNHE}. But we should keep
in mind that for any realistic application, such as in the determination
of the equation of state relevant for the physics of compact stellar 
objects, both of the conditions of color -- electric charge neutrality and 
$\beta$-equilibrium -- should be imposed.

%%%%%%%%%%%%%%%%%%%%%%%%%%%%%%%%%%%%%%%%%%%%%%%%%%%%%%%%%%
\subsection{Numerical results: Absence of color neutrality}

We now turn to the numerical results obtained with the OPT method and
the comparison of these results with those obtained using the LN
approximation.  {}For simplicity and making easier the comparison
between the OPT and LN results, we will first analyze the case of the
absence of color neutrality (e.g., we consider $\mu_8=0$ initially),
and we can assume simply, as previously stated, that 
$\mu_{r}=\mu_{b}=\mu_g\equiv \mu_{B}/3$.

{}From the OPT thermodynamic potential at zero temperature,
$V_{\rm eff}^{OPT}(\sigma_{c},\Delta_{c})$, given by the sum of
Eqs.~(\ref{VeffMFOPT'T=0}), (\ref{Veff2loopq12T=0}), and
(\ref{Veff2loopq3T=0}), together with the corresponding gap equations
for the chiral and diquark  condensates,

\begin{eqnarray} \label{gapNJLsigDelt2}
\left.\frac{\partial V_{\rm eff}}{\partial
  \sigma}\right|_{\sigma=\sigma_{c}} =0,  \qquad \left.\frac{\partial
  V_{\rm eff}}{\partial \Delta}\right|_ {\Delta=\Delta_{c}}=0, 
\end{eqnarray}
and the PMS conditions,  Eqs.~(\ref{pmseta}) and (\ref{pmsalpha}), applied to the OPT mass
parameters $\eta$ and $\alpha$,  we can find numerically the behavior
for the chiral condensate  $\sigma$ (and consequently, that for the
effective quark mass $M_{q}$) and  diquark condensate $\Delta$,  as
well as all the relevant thermodynamic properties of the system, as a
function  of the chemical potential. (For convenience we drop the
subscript $c$ in $\sigma$ and $\Delta$ from now on.)  In the
absence of color neutrality, we can write the effective chemical
potential characterizing the BEC-BCS crossover simply as
$\mu_{N}=\mu_{B}/3-M_{q}$.  As in conventional in the literature, we will
present the results as a  function of the baryon chemical potential
$\mu_{B}$.

%%%%%%%%%%%%%%%%%%%%%%%%%%%%%%%%%%%%%%%%%%%%%%%%%%%%%%%%%%%
\begin{figure}[!htpb]
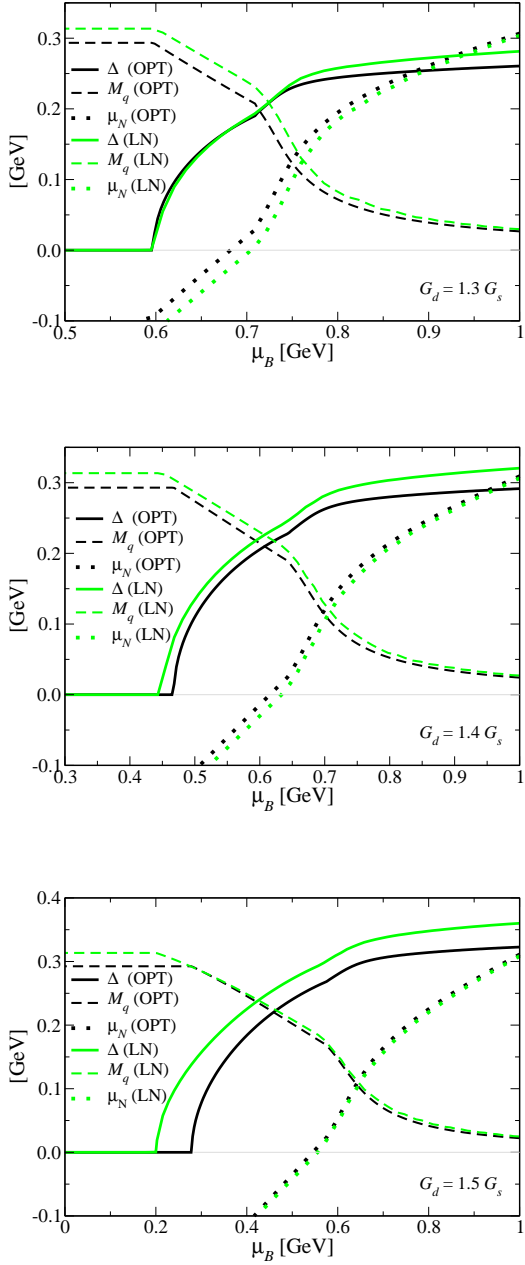

\vspace{0.5cm}
\centerline{ \includegraphics[scale=0.28]{fig5a.eps}}
\vspace{1cm} \centerline{
  \includegraphics[scale=0.28]{fig5b.eps}}
\vspace{1cm} \centerline{
  \includegraphics[scale=0.28]{fig5c.eps}}
\caption{ Diquark condensate $\Delta$, effective  mass $M_{q}$, and
  effective chemical potential $\mu_{N}$ as a function  of the baryon
  chemical potential $\mu_{B}$ for different values of the ratio
  $G_{d}/G_s$ in the LN approximation and OPT comparatively.}
\label{fig: MFOPTsn}
\end{figure}
%%%%%%%%%%%%%%%%%%%%%%%%%%%%%%%%%%%%%%%%%%%%%%%%%%%%%%%%%%%%%%

\begin{figure}[!htb]
\vspace{0.5cm}
\centerline{\includegraphics[scale=0.28]{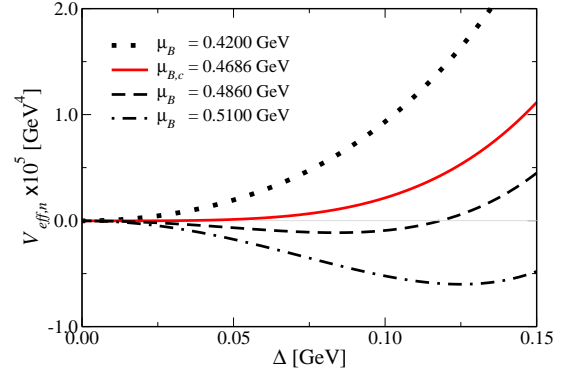}}
\caption{ The effective potential in the OPT at zero  temperature with
  the vacuum energy subtracted, $V_{\rm eff}\equiv
  V_{\rm eff}(\sigma,\Delta,\bar{\eta},\bar{\alpha},\mu_{B})-
  V_{\rm eff}(\sigma_{vac},\Delta=0,\bar{\eta}_{vac},\bar{\alpha},\mu_{B}=0)$,
  as a  function of  $\Delta$ for different values of baryon chemical
  potential  $\mu_{B}$ around the critical value $\mu_{B,c}^{\rm}=
  0.4686$ GeV , for  the ratio $G_{d}/G_{s}=1.4$. The evolution of the
  global mininum of potential  when changing  $\mu_{B}$ suggests a
  second-order phase transition in the order parameter  $\Delta$, as
  occurs in the LN case. }
\label{fig: PotOPTsn14}
\end{figure}

We start by showing in {}Fig.~\ref{fig: MFOPTsn} the behavior of the
effective quark mass $M_{q}$ and $\Delta$ with the increase of  the
baryon chemical potential $\mu_{B}$ for $G_d/G_s=1.3,\, 1.4$ and
$1.5$, which were the same values considered in Ref.~\cite{SUNHE}, which
studied the BEC-BCS crossover in
the LN approximation.  The $\mu_{N}$ result
for each method is also indicated in the plots, such as to facilitate
visualization of the BEC region, which corresponds to the values of
$\mu_{B}$  for which $\mu_{N}<0$, when $\Delta\neq 0$, going to
$\mu_{N}>0$, corresponding to the BCS region.  The results in
{}Fig.~\ref{fig: MFOPTsn} indicate that OPT disfavors the BEC region
and that this region  seems to decrease more significantly with the
decrease of the ratio $G_d/G_s$.  In addition, we  observe that the
OPT also disfavors the region in which $\Delta\neq 0$ for
$1.3\lesssim G_{d}/G_{s}\lesssim 1.5$, increasing the value of
critical chemical  potential(s) $\mu_{B,c}^{\rm}$ (OPT) relative
to those of the LN approximation.  {}From the qualitative point of
view, the variation of $\Delta$ and $M_{q}^{OPT}$  with the variation
of $\mu_{B}$ remain similar to the ones observed in the LN case, while
maintaining the phase transition as being second order when color
neutrality  is not required, as shown in {}Fig.~\ref{fig: PotOPTsn14}.

%%%%%%%%%%%%%%%%%%%%%%%%%%%%%%%%%%%%%%%%%%%%%%%%%%%%%%%
%%%%%%%%%%%%%%%%%%%%%%%%%%%%%%%%%%%%%%%%%%%%%%%%%%%%%%%%%%

By looking again at {}Fig.~\ref{fig: MFOPTsn}, we note that the value
of  the condensate $\Delta$ given by the OPT is always smaller than
the one given  by the LN approximation, and this difference becomes
larger with the increase of  $G_{d}/G_{s}$. In addition, we observe
that $M_{q}^{OPT}$ increasingly  approaches $M_{q}^{LN}$ for
increasing values of $G_{d}/G_{s}$ and $\mu_{B}$.  The difference
between these quantities -- for example, for the case $G_{d}/G_{s}=1.5$ -- is 
visually insignificant from the critical chemical potential
$\mu_{B,c}^{\rm}$ (OPT) of  the phase transition in OPT. Something
similar occurs with the chemical  potential for which occurs the
BEC-BCS crossover, obtained by the condition $\mu_{N}=0$,
$\mu_{B,c}^{\rm BEC-BCS}$. In the OPT, the crossover requires a value
of  $\mu_{B,c}^{\rm BEC-BCS}$ (OPT) that is lower  than that of in the
LN approximation case, and this difference tends to decrease
appreciably with the increase of $G_{d}/G_{s}$.

These results concerning the BEC-BCS crossover and the differences between
the LN and OPT critical values are summarized in the Table~\ref{tab:muc2},
where, for completeness, we also show the value for the pseudocritical
chemical potential, $\mu_{B,pc}^{\rm ch}$, for the chiral symmetry 
crossover (defined by the position
of the inflection point in $M_q$).

%%%%%%%%%%%%%%%%%%%%%%%%%%%%%%%%%%%%%%%%%%%%%%%%%%%%%%%
\begin{table}[!htb]
\caption{Values of critical chemical potentials for
the LN and OPT, in both cases in the absence of color 
neutrality.}
\label{tab:muc2}
\begin{tabular}{ccccc}
\hline  
\multicolumn{5}{c}{No color neutrality case}\\ \hline
 & $G_d / G_s$ & $\mu_{B,c}$ (GeV) & $\mu_{B,c}^{\rm BEC-BCS}$ (GeV)
 & $\mu_{B,pc}^{\rm ch}$ (GeV) \\ \hline 
    &  1.3  & 0.6003  &  0.7051  &  0.7398   \\  
 LN &  1.4  & 0.4513  &  0.6334  &  0.6785   \\ 
    &  1.5  & 0.2010  &  0.5557  &  0.6104   \\ \hline
    &  1.3  & 0.5972  &  0.6820  &  0.7306   \\ 
 OPT&  1.4  & 0.4686  &  0.6155  &  0.6742   \\ 
    &  1.5  & 0.2787  &  0.5454  &  0.6134   \\ \hline
\end{tabular}
\end{table}
%%%%%%%%%%%%%%%%%%%%%%%%%%%%%%%%%%%%%%%%%%%%%%%%%%%%%%

%%%%%%%%%%%%%%%%%%%%%%%%%%%%%%%%%%%%%%%%%%%%%%%%%%%%%%%%%%
\begin{figure}[!htb]
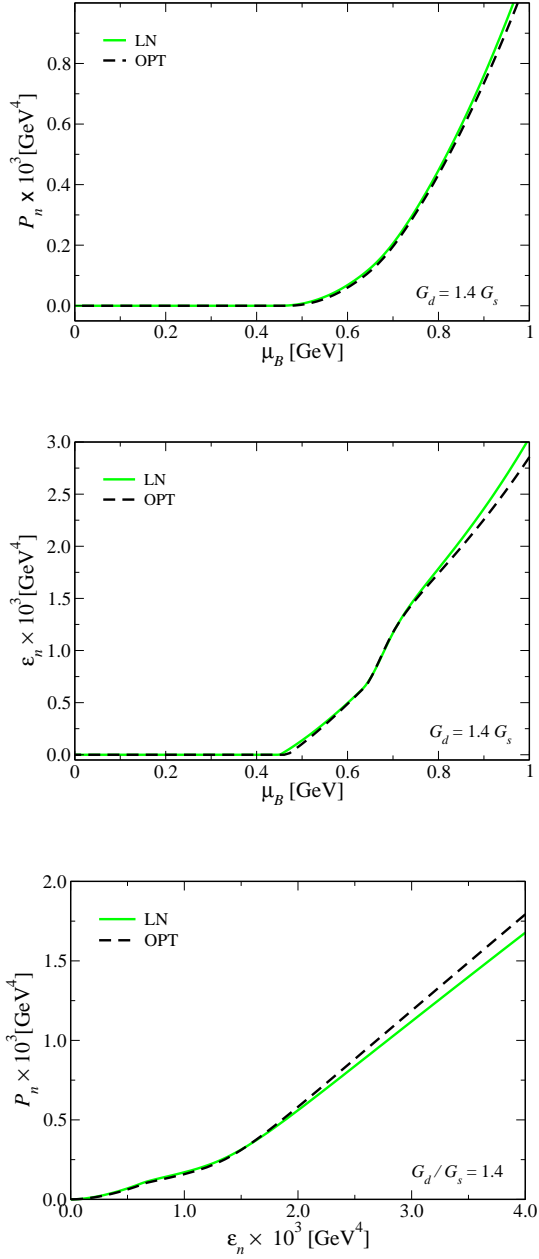

\vspace{0.5cm}
\centerline{ \includegraphics[scale=0.28]{fig7a.eps}}
\vspace{0.85cm} \centerline{
  \includegraphics[scale=0.28]{fig7b.eps}}
\vspace{0.85cm} \centerline{
  \includegraphics[scale=0.28]{fig7c.eps}}
\caption{ The pressure (top plot), the energy density as a
  function of $\mu_B$ (middle plot) and the equation of state (bottom plot), both
  for the ratio $G_{d}/G_{s}=1.4$.}
\label{fig: MFOPT1.4sn2}
\end{figure}
%%%%%%%%%%%%%%%%%%%%%%%%%%%%%%%%%%%%%%%%%%%%%%%%%%%%%%%%%

To also exemplify some of the differences between OPT and LN for other
thermodynamic quantities, in {}Fig.~\ref{fig: MFOPT1.4sn2} we show the
vacuum subtracted  pressure and energy densities,
$P_{n}(\sigma,\Delta)$ and  $\varepsilon_{n}(\sigma,\Delta)$, respectively, in
addition to the  equation of state $P_{n}(\varepsilon_{n})$, where
$P_n=P-P_{vac}$ and $\varepsilon_n=\varepsilon-\varepsilon_{vac}$, with
(at $T=0$)

\begin{eqnarray} \label{defPnEn}
P(\sigma,\Delta)&=&-V_{\rm eff}(\sigma,\Delta),
\\ \varepsilon(\sigma,\Delta)&=&-P(\sigma,\Delta)+\mu_B n_{B},
\end{eqnarray}  
where $n_{B}$ is the baryon number density, given by 

\begin{eqnarray}
n_{B}=-\frac{\partial}{\partial\mu_{B}}V_{\rm eff}(\sigma,\Delta).
\end{eqnarray}

We have restricted {}Fig.~\ref{fig: MFOPT1.4sn2} to show only  the
case $G_{d}/G_{s}=1.4$ as an example.  Visually, there is no
significant differences between such cases in the region of interest.
But in the region of intermediate baryon chemical potentials,  the OPT
slightly decreases its values compared to the LN approximation for the
value  of $G_{d}/G_{s}$ used herein.

%%%%%%%%%%%%%%%%%%%%%%%%%%%%%%%%%%%%%%%%%%%%%%%%%%%%%%%%%%%%%%%%%%%%%
\subsection{The OPT results in the case of color neutrality}
\label{OPTneutral}

Let us now consider the case of imposing the color neutrality
condition.  As already discussed above, in this case we set
$\mu=\mu_{B}/3+\mu_{8}/3$, $\mu_{b}=\mu_{B}/3-2\mu_{8}/3$,  and the
condition of color neutrality, given by
Eq.~(\ref{neutralitycondition}),  must be satisfied in  the region
where $\Delta\neq 0$, that represents the physical case.

%%%%%%%%%%%%%%%%%%%%%%%%%%%%%%%%%%%%%%%%%%%%%%%%%%%%%%
\begin{figure}[!htb]
\vspace{0.5cm}
\centerline{ \includegraphics[scale=0.28]{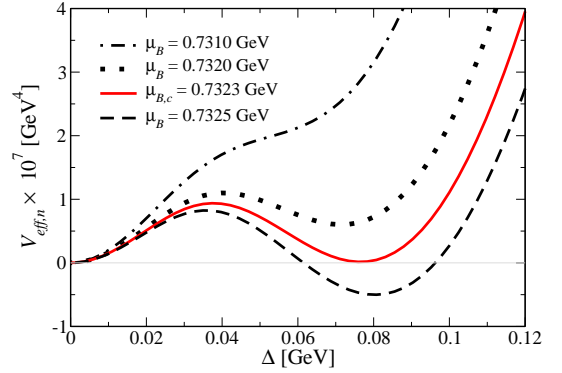}}
\caption{ The vacuum subtracted effective potential in the OPT at zero
  temperature as a  function of  $\Delta$, for different values of the
  baryon chemical potential  $\mu_{B}$ around the critical value
  $\mu_{B,c} = 0.7323$ GeV and for the ratio $G_{d}/G_{s}=1.3$. The
  behavior of the global minimum of potential with the variation of
  $\mu_{B}$ and the coexistence between the two minima suggests  a
  first-order phase transition in the order parameter $\Delta$.}
\label{fig: PotOPTcn13}
\end{figure}
%%%%%%%%%%%%%%%%%%%%%%%%%%%%%%%%%%%%%%%%%%%%%%%%%%%%%

%%%%%%%%%%%%%%%FIGURE%%%%%%%%%%%%%%%%%%%
\begin{center}
\begin{figure}[!htb]
\vspace{0.5cm}
{\includegraphics[scale=0.28]{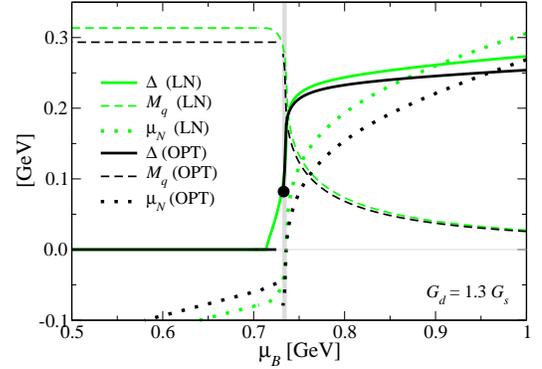}}
\caption{The diquark condensate $\Delta$, the effective quark mass
  $M_{q}$, and the effective chemical potential $\mu_{N}$ as a function
  of the baryon chemical potential $\mu_{B}$ for $G_{d}/G_{s} = 1.3$,
  in  the LN approximation and OPT comparatively.  Color
  neutrality is considered. The thin vertical line indicates  the
position of the first-order transition (discontinuity) in $\Delta$.}
\label{fig: muN1p3neut}
\end{figure}
\end{center}
%%%%%%%%%%%%%%%%%%%%%%%%%%%%%%%%%%%%%%%%

The main effect coming from the corrections due to the OPT in relation
to the case of LN,  for the values of $G_{d}/G_{s}$ previously
considered, is that there is a  discontinuity in
$\Delta(\mu_{B})$ at the critical baryon  chemical potential
$\mu_{B,c}$ (OPT), indicating a first-order phase transition.  The
emergence of a first-order transition in this case can be confirmed
and  illustrated in {}Figs.~\ref{fig: PotOPTcn13}  and~\ref{fig:
  muN1p3neut}, where in both cases we have considered the case
$G_{d}/G_{s}=1.3$ as an example.  
In this case, when  the baryon chemical potential increases, the potential
presents a new (local) minimum around  $\mu_B \simeq 0.7113$ GeV, and
at the critical baryon chemical potential $\mu_{B,c} \simeq 0.73226$ 
GeV, this minimum is aligned to the one at  $\Delta = 0$. If we keep
increasing the chemical potential, the minimum at  origin becomes
local, and after that, a maximum point, around $\mu_B\simeq 0.7365$
GeV emerges.  This interval, 0.7113 GeV $\lesssim \mu_B \lesssim $
0.7365 GeV, corresponds to  a metastable region, represented by the
thin vertical gray region in   {}Fig.~\ref{fig: muN1p3neut}. 
In the LN case, the BEC region, when contrasted with the case shown
in {}Fig.~\ref{fig: MFOPTsn} obtained when neglecting color neutrality,
also shrinks, but it does not disappear completely, consistent with
the observations made in Ref.~\cite{SUNHE}.

We should remark that in the LN approximation, the transition eventually 
also turns first order, but for values of the
ratio $G_d/G_s \lesssim 1.05$, as observed in
Refs.~\cite{SUNHE,Kitazawa:2007zs}.
By increasing the ratio  $G_d/G_s$, we can again
recover a second-order transition for the phase with a diquark
condensate and a BEC-BCS crossover. {}For the OPT case, we find that the
minimum value required for a second-order phase transition shifts from
$G_d/G_s \approx 1.05$ in the LN case to a value $G_d/G_s \approx
1.525$, which is itself very close to the maximum value allowed for
the ratio $G_d/G_s$ before the mass of the diquark vanishes,
precluding the instability of the vacuum. In the OPT, for the
parameters considered, this happens for values   $G_d/G_s > 1.55$.

%%%%%%%%%%%%%%%%%%%%%%%%%%%%%%%%%%%%%%%%%%%%%%%%%%%%%%%
\begin{figure}[!htb]
\vspace{0.5cm}
\centerline{ \includegraphics[scale=0.28]{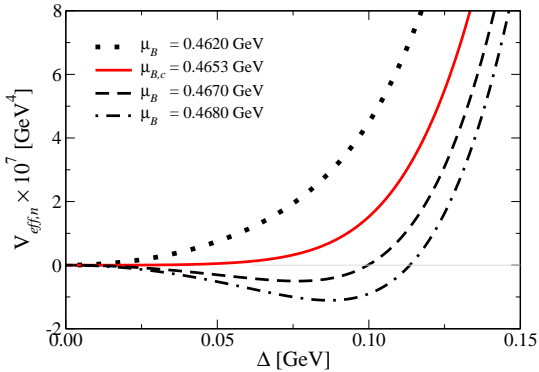}}
\caption{ The vacuum subtracted effective potential in the OPT at zero
  temperature as a  function of  $\Delta$ for different values of
  baryon chemical potential  $\mu_{B}$ around the critical value
  $\mu_{B,c}^{\rm} = 0.4653$ GeV, for  $G_{d}/G_{s}=1.53$. The
  behavior of the global minimum of potential with the variation  of
  $\mu_{B}$ suggests a second-order phase transition in the order
  parameter  $\Delta$. Color neutrality is considered.}
\label{fig: PotOPTsn153}
\end{figure}
%%%%%%%%%%%%%%%%%%%%%%%%%%%%%%%%%%%%%%%%%%%%%%%%%%%%%%%%%%

In {}Fig.~\ref{fig: PotOPTsn153}, we show the effective potential in
the OPT for the case of  $G_d/G_s=1.53$, confirming the resurgence of
the second-order phase transition for diquark condensation.

Next, we will restrict our attention to the cases where a second-order phase
transition for diquark condensation is possible in the OPT, which will
in particular correspond  to the cases where the ratio of $G_d/G_s$
will assume the values $G_d/G_s=1.53,\, 1.54$, and $1.55$.

\begin{figure}[!htb]
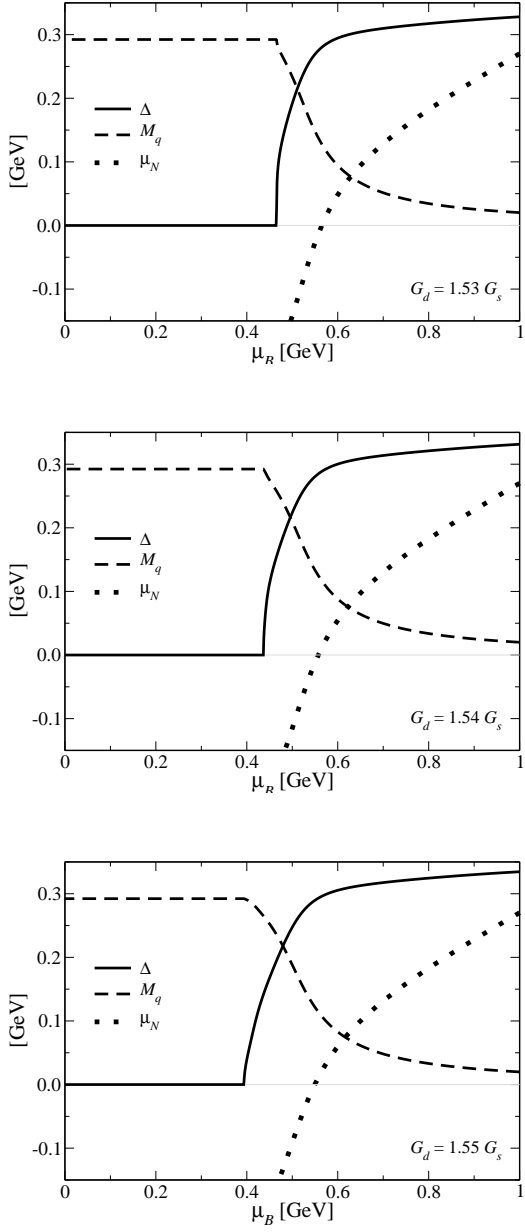

\vspace{0.5cm}
\centerline{ \includegraphics[scale=0.28]{fig11a.eps}}
\vspace{0.8cm} \centerline{
  \includegraphics[scale=0.28]{fig11b.eps}}
\vspace{0.8cm} \centerline{
  \includegraphics[scale=0.28]{fig11c.eps}}
\caption{ Diquark condensate $\Delta$, effective  mass $M_{q}$, and
  effective chemical potential $\mu_{N}$ as a function  of baryon
  chemical potential $\mu_{B}$ for different values of $G_{d}/G_{s}$
  in the OPT case.  Color neutrality is considered.}
\label{fig: compOPTcn}
\end{figure}

\begin{center}
\begin{figure*}[!htb]
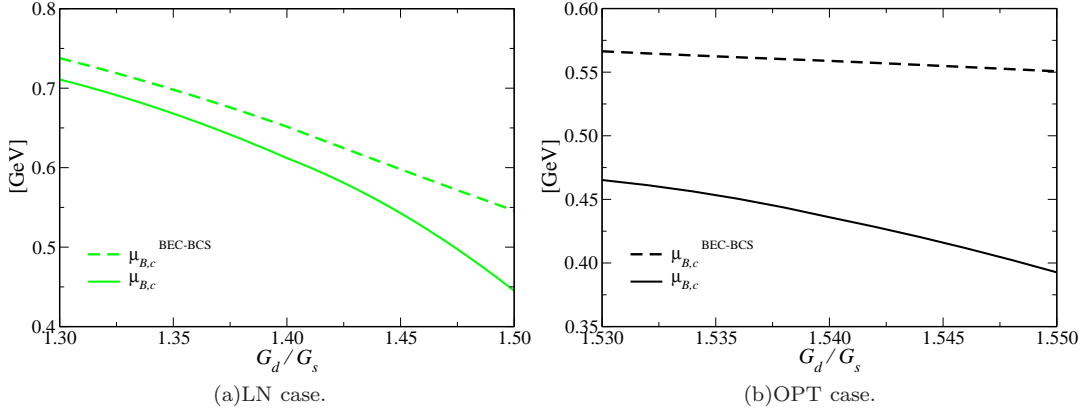

\vspace{0.5cm}
\subfigure[LN case.]{\includegraphics[scale=0.28]{fig12a.eps}}
\subfigure[OPT case.]{\includegraphics[scale=0.28]{fig12b.eps}}
\caption{The critical chemical potentials  associated with the BEC
  phase transition $\mu_{B,c}^{\rm}$,  as a function
  of $G_{d}/G_{s}$, for the LN and OPT cases, in their correspondent
  validity range. Color neutrality is considered in both cases.}
\label{fig: BECreg}
\end{figure*}
\end{center}

In {}Fig.~\ref{fig: compOPTcn}, we show the results for $\Delta, M_q$,
and $\mu_N$ for  the values $G_d/G_s=1.53,\, 1.54$, and $1.55$. It is
possible to see that, similarly to the  LN case with results shown in
 {}Fig.~\ref{fig: MFOPTsn}, as we increase the ratio  $G_d/G_s$, the OPT
favors the BEC phase. The critical baryon  chemical potential $\mu_{B,c}^{\rm}$
and the crossover value
$\mu_{B,c}^{\rm BEC-BCS}$ both decrease as the ratio $G_d/G_s$ increases.
But $\mu_{B,c}^{\rm}$ is more affected by the value of $G_d/G_s$.
In the LN case, however, both $\mu_{B,c}^{\rm}$ and $\mu_{B,c}^{\rm BEC-BCS}$
suffer similar influence due to a variation of of $G_d/G_s$. 
Both of these results can be seen in
{}Fig.~\ref{fig: BECreg}.

%%%%%%%%%%%%%%%%%%%%%%%%%%%%%%%%%%%%%%%%%%%%%%%%%%%%%%%%
%%%%%%%%%%%%%%%%%%%%%%%%%%%%%%%%%%%%%%%%%%%%%%%%%%%%%%%

\begin{center}
\begin{figure*}[!htb]
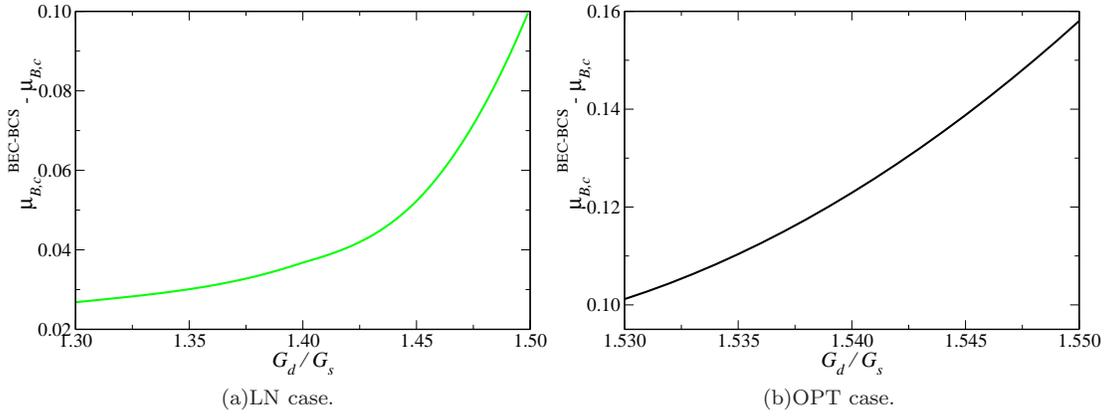

\vspace{0.5cm}
\subfigure[LN  case.]{\includegraphics[scale=0.28]{fig13a.eps}} 
\subfigure[OPT  case.]{\includegraphics[scale=0.28]{fig13b.eps}}
\caption{The width of the BEC region, defined
  by $(\mu_{B,c}^{\rm BEC-BCS}-\mu_{B,c}^{\rm})$,  as a function
  of $G_{d}/G_{s}$, for the LN and OPT cases, in their correspondent
  validity range. Color neutrality is considered in both cases.}
\label{fig: BECreg-width}
\end{figure*}
\end{center}

In {}Fig.~\ref{fig: BECreg}, we illustrate the evolution of the
critical points  $\mu_{B,c}^{\rm}$, $\mu_{B,c}^{\rm BEC-BCS}$,
while in {}Fig.~\ref{fig: BECreg-width} we give
the width of the BEC region, defined by $(\mu_{B,c}^{\rm
  BEC-BCS}-\mu_{B,c}^{\rm})$ as  a function of $G_{d}/G_{s}$, for
both the LN and OPT cases.

%%%%%%%%%%%%%%%FIGURE%%%%%%%%%%%%%%%%%%%

%%%%%%%%%%%%%%%FIGURE%%%%%%%%%%%%%%%%%%%
%%%%%%%%%%%%%%%%%%%%%%%%%%%%%%%%%%%%%%%%

In Table~\ref{tab:muc-color} we summarize the values for the critical chemical potentials 
obtained when considering the color neutrality condition
in the LN and OPT cases. {}For completeness, we also give the values
for the pseudocritical chemical potential for chiral condensation, $\mu_{B,pc}^{\rm ch}$. 
Note that the critical baryon  chemical potential for the BEC transition 
always tends to decrease as we increase the ratio $G_d / G_s$, which is true in both the
LN and OPT cases. Note also that the results for the critical baryon  chemical potentials 
always remain below that of the value of the onset of baryonic matter (e.g., when
comparing with the nucleon mass), which 
prompts the question of the reliability of these results when applied to real
QCD. In fact, the same trend we see here is also seen in all previous studies 
for the BEC-BCS crossover study in the NJL model (see, however, Ref.~\cite{He:2010nb}).
As far as this issue is concerned, when we compare the LN and OPT results, we see that
while the LN gives a much larger range of values for $G_d / G_s$ allowing for the BEC
phase, in the OPT case this window shrinks considerably to a very small
range of values, $1.525 \lesssim G_d/G_s \lesssim 1.55$. In a sense, by including further
contributions from both meson and diquark fluctuations (represented by the 
two-loop contributions) which are absent in the LN approximation, the OPT
clearly disfavors the emergence of a BEC phase. This seems more in accordance,
based on these results, with the expectancy that
the appearance of a diquark BEC phase at low density must be an artificial effect in the 
(three-color) NJL model.

%%%%%%%%%%%%%%%%%%%%%%%%%%%%%%%%%%%%%%%%%%%%%%%%%%%%%%
\begin{table}[!htb]
\caption{Values of critical chemical potential, considering color neutrality 
effects, for LN and OPT.}
\label{tab:muc-color}
\begin{tabular}{ccccc}
\hline  
\multicolumn{5}{c}{The color neutrality case}\\ \hline    
 & $G_d / G_s$ & $\mu_{B,c}$ (GeV) & $\mu_{B,c}^{\rm BEC-BCS}$ (GeV)
 & $\mu_{B,pc}^{\rm ch}$ (GeV) \\ \hline 
    &  1.3  &  0.7137 &  0.7370  &  0.7361   \\ 
 LN &  1.4  &  0.6144 &  0.6603  &  0.6459   \\ 
    &  1.5  &  0.4474 &  0.5767  &  0.5334   \\ \hline
\centering{\multirow{4}{0.8cm}{OPT}}     
    & 1.53  &  0.4653 &  0.5651  &  0.5213   \\ 
    & 1.54  &  0.4366 &  0.5573  &  0.5119   \\ 
    & 1.55  &  0.3939 &  0.5496  &  0.5025   \\ \hline 
\end{tabular}
\end{table}
%%%%%%%%%%%%%%%%%%%%%%%%%%%%%%%%%%%%%%%%%%%%%%%%%%%%%%%%

%%%%%%%%%%%%%%%%%%%%%%%%%%%%%%%%%%%%%%%%%%%%%%%%%%%%%%%%%%%%%%%%%%%%%%%%%%
\begin{figure}[!htb]
\vspace{0.5cm}
\centerline{ \includegraphics[scale=0.28]{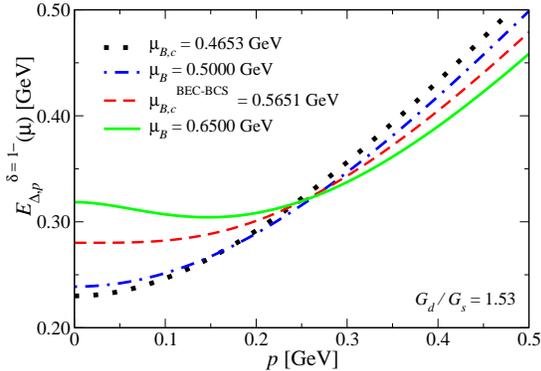}}
\caption{Particle dispersion relation $E_{\Delta,\vec{p}}^{\delta=1-}(\mu)$ in the
  OPT approximation for $G_d/G_s = 1.53$.  Color neutrality is
  considered.}
\label{dispOPT}
\end{figure}
%%%%%%%%%%%%%%%%%%%%%%%%%%%%%%%%%%%%%%%%%%%%%%%%%%%%%%%%%%%%%%%%%%%%%%%%

As already mentioned previously, the BEC-BCS crossover can be
characterized  by the shape of the dispersion relation for the quark
field, which for the OPT case is given by  Eq.~(\ref{fermion-E}) when
setting  $\delta = 1, \Delta_1 = \Delta_2=\Delta$, $\alpha_1 =
\alpha_2=\alpha$, and $\mu'=\mu=\mu_{B}/3+\mu_{8}/3$, or $E_{\Delta,\vec{p}}^{\delta=1-}(\mu)
=\sqrt{\left[\sqrt{p^2+(M_q^{OPT})^2}-\mu\right]^{2}
  +\Delta^2}$ (the same form as in the LN). In
{}Fig.~\ref{dispOPT}, we  illustrate the particle dispersion for the OPT
in the color neutrality case for the example of $G_d/G_s=1.53$. {}For
values of $\mu_B\leq \mu_{B,c}^{\rm BEC-BCS}$, the  minimum of the dispersion is located at
$\vec{p} = 0$, and the gap energy is $|\mu_N|$ (remembering that
$\mu_N = \mu - M_q^{OPT}$, where $M_q^{OPT} = m + \sigma^{OPT}$).  
The BEC phase corresponds to the region between $\mu_{B,c}$ and $\mu_{B,c}^{\rm BEC-BCS}$.
Increasing the
chemical potential beyond $\mu_{B,c}^{\rm BEC-BCS}$, this minimum is shifted to 
$|\vec{p}|\simeq |\mu|$
and the gap  becomes equal to $\Delta$, indicating the BCS phase.
Note that the diquark condensate $\Delta$, as can be seen from {}Figs.~\ref{fig: MFOPTsn}
and \ref{fig: compOPTcn}, tends to remain smaller than the baryon chemical potential.
We also find that the critical chemical potential for the BEC transition 
($\mu_c\equiv \mu_{B,c}/3$) corresponds exactly to half the mass of the diquarks. 
This can be proofed as follows: The diquark mass can be computed in the OPT
scheme similarly to the calculation of the pion mass
shown in Sec.~\ref{pionmass:Sec}, with the appropriate changes -- e.g., by replacing
the pion vertex $-i \gamma^5 \tau_i$ with that of the diquark
boson field with the quarks, $iC\gamma^5 \tau_2 t_2$ obtained from the
bosonized Lagrangian density Eq.~(\ref{LB}) -- from which we then obtain that the
diquark mass is determined by the pole equation obtained in terms of
the diquark self-energy $\Pi_d(q_0,\vec q\,)$ as

\begin{eqnarray} 
0&=&1-2G_d \Pi_d(q_0=m_d,\vec q=0)
\nonumber \\
&=&1-2G_{d}\left\{  2i(N_{c}-1)N_{f}\left[2 I_G-m_{d}^2I(m_{d}^2)\right] 
\right. \nonumber\\
&-&\left. 8(N_c-1)N_f G_{s}\left[2I_G-m_d^2 I(m_d^2)
\right]^2 \right. \nonumber\\
&+& \left. 16(N_c-1)N_f G_s m_d^2 \mathcal{M}^2 I^2(m_d^2)          \right\},
\label{diquarkmass}
\end{eqnarray}
where $I_G$ is given by Eq.~(\ref{IG}) and $I(m_d)$ is obtained from Eq.~(\ref{I(q2)}).
Note that Eq.~(\ref{diquarkmass}) in the LN limit reduces to

\begin{eqnarray} 
1&=&4iG_{d}(N_{c}-1)N_{f}\left[2 I_G-m_{d}^2I(m_{d}^2)\right]
\nonumber \\
&=&2G_{d}(N_{c}-1)N_{f}\int \frac{d^3p}{(2 \pi)^3}
\left( \frac{1}{E_{\vec p}+m_d/2} 
\right.
\nonumber \\
&+& \left.  \frac{1}{E_{\vec p}-m_d/2}\right),
\label{diquarkmassLN}
\end{eqnarray}
where we have evaluated the integral in $p_0$ to obtain the last line in
the above equation. Equation~(\ref{diquarkmassLN}) agrees with the corresponding
LN result of Refs.~\cite{SUNHE,ZHUANG}.
In the OPT case, Eq.~(\ref{diquarkmass}) is a function of the optimization parameter
$\eta$ and must then be solved together with the PMS equation (\ref{pmseta}).
Equation~(\ref{diquarkmassLN}) in the LN approximation can be compared with the
one determining the diquark condensate $\Delta$ [Eq.~(\ref{gapdelta})], and
use of the PMS equations (\ref{pmseta}) and (\ref{pmsalpha}), gives

\begin{equation}
\Delta = 2(N_{c}-1)N_{f} G_d  \bar{\alpha} J_B(\mu)\Bigr|_{\eta={\bar \eta}},
\label{eqDelta}
\end{equation}
where $J_B(\mu)$ is given by Eq.~(\ref{JB}). If we set again the LN limit in 
Eq.~(\ref{eqDelta}) and recall that in this case
the OPT optimization parameters $\bar \eta$ and $\bar \alpha$ reduce to 
$\bar \eta = \sigma$ and $\bar \alpha = \Delta$, respectively,
then Eq.~(\ref{eqDelta})
becomes, at the diquark condensation point $\mu_B \to \mu_{B,c}$
and where $\Delta\to 0$,

\begin{eqnarray}
1 &=& 2(N_{c}-1)N_{f} G_d  \int \frac{d^3p}{(2 \pi)^3}
\left( \frac{1}{E_{\vec p}+\mu_{B,c}/3} \right.
\nonumber \\
&+& \left.   \frac{1}{E_{\vec p}-\mu_{B,c}/3}\right),
\label{eqDeltaLN}
\end{eqnarray}
and we also recover the LN result for $\mu_{B,c}$ as given in Refs.~\cite{SUNHE,ZHUANG}.
When comparing Eq.~(\ref{diquarkmassLN}) with Eq.~(\ref{eqDeltaLN}),
we see immediately that $\mu_{B,c}/3=m_d/2$.
Note also that when accounting for color neutrality, the same result follows when
we consider that $\mu\to \mu + \mu_8/3$ [note that the integral $J_B$ in 
Eq.~(\ref{eqDelta}) is only a function of $\mu=\mu_r=\mu_g$], and the critical
baryon chemical potential for diquark condensation shifts accordingly,
$\mu_{B,c}/3 \to m_d/2- \mu_8/3$. Since $\mu_8$ is in general negative, this corresponds
to a increase of the diquark condensation point when color neutrality is considered,
which agrees with the results shown, e.g., in Table~\ref{tab:muc-color}. 
In the OPT scheme, this comparison
between the diquark mass and the value for the condensation point is more 
involved for two main reasons: 
{}First, because now we have to solve Eqs.~(\ref{diquarkmass}) and
(\ref{eqDelta}) subject to the PMS Eqs.~(\ref{pmseta}) and (\ref{pmsalpha})
and also the gap equation determining $M_q$, which makes the numerical work
somewhat more involved. Second and most importantly, there is clearly
a mismatch between the order-1 OPT expression for the diquark self-energy
leading to Eq.~(\ref{diquarkmass}) and the corresponding contributions considered at
the same order in the OPT for the effective potential. In particular, note that
the order-1 OPT contributions to the diquark mass, corresponding to the two-loop
diagrams which are similar to the ones seen in {}Fig.~\ref{fig: FDparameters}
for the pion, turn out to be equivalent to three-loop vacuum diagrams in the effective
potential (e.g., when we close the external diquark legs in the self-energy
diagrams such as to construct equivalent vacuum terms). These contributions
would in fact be order2 in the OPT scheme when seen in the context of the effective potential. 
Due to this mismatch of terms between the diquark self-energy and the effective potential,
we do not expect a perfect agreement for the value of $\mu_{B,c}/3$ obtained
from the optimization of the effective potential, with the value of $m_d/2$ obtained
from Eq.~(\ref{diquarkmass}). This is a feature of the OPT scheme. Intrinsically,
we should optimize a quantity that could produce simultaneous values for both
$\mu_{B,c}$ and $m_d$. This could be, perhaps, the nonhomogeneous (space- or 
momentum-dependent) effective action, instead of the effective potential (the zero-momentum 
homogeneous action). Even so, when we compare results from these different
quantities, we obtain, taking as an example the ratio $G_d/G_s=1.4$ in the absence
of charge neutrality, the result $\mu_{B,c}/3=0.1562$ GeV, while the result from 
Eq.~(\ref{diquarkmass}) gives $m_d/2 = 0.1596$ GeV, a difference of around $2\%$.
Though not a proof, we can take this difference as a rough possible
indication of the convergence of the OPT and a signal that when going to the next order, which
will now include three-loop terms with similar topology to the ones contributing
at the self-energy for the diquark mass, these terms are expected to produce an
overall small contribution.
This is a generic expectation from the OPT scheme seen in studies of
its convergence properties in other models~\cite{opt6}.

\begin{figure}[!htb]
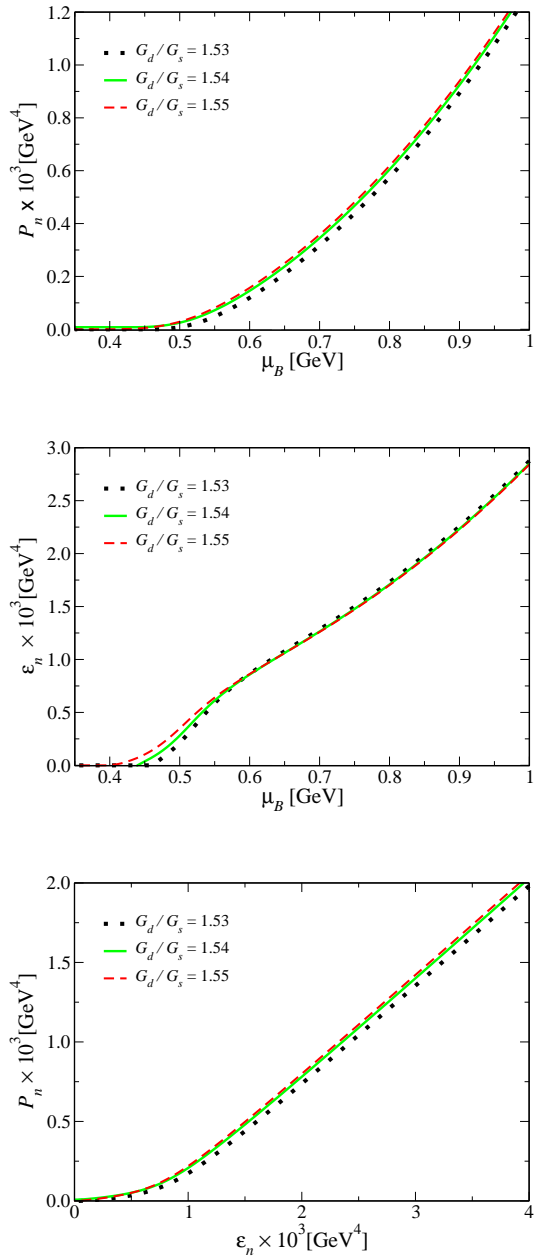

\centerline{ \includegraphics[scale=0.28]{fig15a.eps}}
\vspace{0.8cm} \centerline{ \includegraphics[scale=0.28]{fig15b.eps}}
\vspace{0.8cm} \centerline{
  \includegraphics[scale=0.28]{fig15c.eps}}
\caption{The thermodynamic quantities of the  system (normalized
  pressure $P_{n}$ and normalized energy density $\varepsilon_{n}$)  are given
  for different values of $G_{d}/G_{s}$ as a function of the baryon
  chemical potential $\mu_{B}$ at zero temperature on OPT
  approximation,  as well as the state equation $P_{n}(\varepsilon_{n})$. Color
  neutrality is considered.}
\label{fig: ThOPTcn}
\end{figure}

{}Finally, the observations already made in the absence of color
neutrality regarding the thermodynamic quantities, like the pressure,
energy density, and equation of state, remain essentially  the same for
the case with color neutrality. In {}Fig.~\ref{fig: ThOPTcn}, we show
these quantities for the OPT for the three values of $G_s/G_d$
considered in the color neutrality example.

%%%%%%%%%%%%%%%%%%%%%%%%%%%%%%%%%%%%%%%%%%%%%%%%%%%%%%%%%%%%
%%%%%%%%%%%%%%%%%%%%%%%%%%%%%%%%%%%%%%%%%%%%%%%%%%%%%%%%%%%%%%%%%%%%

%%%%%%%%%%%%%%%%%%%%%%%%%%%%%%%%%%%%%%%%%%%%%%%%%%%%%%%%%%%%%%%%%%%%%
\section{Conclusions}
\label{conclusions}

We have studied the BEC-BCS crossover in an extended two-flavor NJL
model, with three  colors and including the diquark interactions,
in the context of  the nonperturbative OPT, method and the results
obtained were contrasted  with those of the usual LN  approximation.
We derived in detail how the fitting of the parameters changes in
the OPT case, deriving the corresponding corrections due to the OPT
for the pion mass and decay constant. These quantities are affected by
the diquark fluctuations already at first order in the OPT
approximation and must be evaluated consistently. 

We have studied the cases both of without and with color neutrality and
have shown the differences between the two cases.  There is a region of
parameter values corresponding to the ratio between  diquarks and the
usual quark-antiquark interactions, $G_d/G_s$, below which a BEC phase
becomes disfavored and the transition from the chiral phase with no
diquark condensate to the phase of diquark condensate is first order,
while for larger values diquarks become massless, condensing already
at vanishing baryon chemical potential, signaling the instability of
the vacuum.   In the absence of color neutrality, for both the LN and
OPT cases, this corresponds approximately to values of the ratio
$G_d/G_s$ satisfying   $1.05 \lesssim G_d/G_s \lesssim 1.52$. When
accounting for color neutrality, this range of values remains roughly
unaltered in the LN case, but for the OPT and the values of the parameters
considered, it slightly shifts and shrinks to the values $1.525 \lesssim
G_d/G_s \lesssim 1.55$.  This shows that the OPT tends to suppress the
BEC region, and consequently, the BEC-BCS crossover.  To our
knowledge, this is the first time that a method beyond the LN, when
applied to the study of the BEC-BCS crossover, has given an indication
of a possible suppression of the BEC regime.  It would be
interesting to further explore this issue when using other
nonperturbative methods or including additional ingredients in the NJL
Lagrangian density, like asymmetries -- for example, a chiral imbalance and
application of the recent regularization method exposed in
Ref.~\cite{FariasMSS} -- or by including a vector meson interaction, as
studied in the LN context for the BEC-BCS crossover in
Ref.~\cite{SUNHE}. With respect to this, it is interesting to point out
that the OPT is able to radiatively generate  vectorlike
interactions~\cite{Kneur:2012qp,Restrepo:2014fna}, which in principle
could also be combined with other effects and possibly change the
BEC-BCS region in  nontrivial ways, as already indicated by the results
of the present work.

%%%%%%%%%%%%%%%%%%%%%%%%%%%%%%%%%%%%%%%%%%%%%%%%%%%%
\acknowledgments   R.~O.~R.~is partially supported by  Conselho
Nacional de Desenvolvimento Cient\'{\i}fico e Tecnol\'ogico -- CNPq
(Grant No.~303377/2013-5), Funda\c{c}\~ao Carlos Chagas Filho de
Amparo \`a Pesquisa do Estado do Rio de Janeiro -- FAPERJ (Grant No. E-26/201.424/2014) 
and Coordena\c{c}\~ao de Pessoal de N\'{\i}vel Superior
-- CAPES (Processo No.~88881.119017/2016-01). 
R. L. S. F. is partially supported by Conselho Nacional de
Desenvolvimento Cient\'{\i}fico e Tecnol\'ogico -- CNPq under Grant
numbers 475110/2013-7, 232766/2014-2 and 308828/2013-5.
R. L. S. F. is also grateful to Lianyi He for insightful comments.
 
%%%%%%%%%%%%%%%%%%%%%%%%%%%%%%%%%%%%%%%%%%%%%%%%%%%%

\appendix

\section{The pion decay constant derivation in the OPT expansion
to order $\delta$. } 
\label{apendice_decayconst}

At one-loop order, the expression of the pion decay constant is  (in
this appendix, we will use the following
notation for the trace: $\mathrm{Tr}\equiv
\mathrm{Tr}_{c}\mathrm{Tr}_{f}\mathrm{Tr}_{D}$)

\begin{eqnarray}
if_{\pi,1}^{2}g_{\mu\nu}\delta_{ij}&=&\frac{1}{4}\int
\frac{d^{4}p}{(2\pi)^{4}} \nonumber \\ &\times&
\textrm{Tr}\left[\frac{i}{\slashed{p}-\mathcal{M}}(\tau_{i}
\gamma_{\mu}\gamma^{5})
  \frac{i}{\slashed{p}+\slashed{q}-\mathcal{M}}(\tau_{j}
\gamma_{\nu}\gamma^{5})\right]
\nonumber\\ &=&-\frac{N_{c}N_{f}}{4}\delta_{ij} \nonumber\\ 
&\times&
\int
\frac{d^{4}p}{(2\pi)^{4}}\frac{1}{(p^{2}-\mathcal{M}^{2}+i\epsilon)((p+q)^{2}
  -\mathcal{M}^{2}+i\epsilon)}\nonumber\\ &\times&\textrm{Tr}_{D}
\left[(\slashed{p}+\mathcal{M})\gamma_{\mu}\gamma^{5}
  (\slashed{p}+\slashed{q}+\mathcal{M})\gamma_{\nu}\gamma^{5}\right]. 
\nonumber\\
\end{eqnarray}
When $q=0$ (zero external momentum), we obtain

\begin{eqnarray}
f_{\pi,1}^{2}g_{\mu\nu}&=&i\frac{N_{c}N_{f}}{4}\int
\frac{d^{4}p}{(2\pi)^{4}}
\left[8p_{\mu}p_{\nu}-4g_{\mu\nu}(p\cdot
  p+\mathcal{M}^{2})\right]
\nonumber\\ &\times&
\frac{1}{(p^{2}-\mathcal{M}^{2}+i\epsilon)[(p+q)^{2}-\mathcal{M}^{2}+i\epsilon]}
\nonumber\\ 
&=&iN_{c}N_{f}\left[2\int
  \frac{d^{4}p}{(2\pi)^{4}}
  \frac{p_{\mu}p_{\nu}}{(p^{2}-\mathcal{M}^{2}+i\epsilon)^{2}}
  \right. \nonumber\\  &-&
  \left. g_{\mu\nu}I_{G}(\mathcal{M})-2g_{\mu\nu}\mathcal{M}^{2}I(0)
  \vphantom{\int
    \frac{d^{4}p}{(2\pi)^{4}}\frac{p_{\mu}p_{\nu}}{(p^{2}-\mathcal{M}^{2}+i\epsilon)^{2}}}
  \right],
\end{eqnarray}
where

\begin{eqnarray}
I_{G}(\mathcal{M})&=&\int\frac{d^{4}p}{(2\pi)^{4}}\frac{1}{p^{2}-\mathcal{M}^{2}+i\epsilon}, 
\end{eqnarray}
and

\begin{eqnarray}
I(0)&=&\int\frac{d^{4}p}{(2\pi)^{4}}\frac{1}{(p^{2}-\mathcal{M}^{2}+i\epsilon)^{2}}.
\end{eqnarray}
Then, using the relation from dimensional regularization~\cite{Itzy}

\begin{eqnarray}
\int
\frac{d^{4}p}{(2\pi)^{4}}\frac{p_{\mu}p_{\nu}}{(p^{2}-\mathcal{M}^{2}+i\epsilon)^{2}}
=\frac{g_{\mu\nu}}{2}I_G(\mathcal{M}), \label{DimReg}
\end{eqnarray}
we obtain~\cite{RUDOPT}

\begin{eqnarray}
f_{\pi,1}^{2}&=&-2iN_{c}N_{f}\mathcal{M}^{2}I(0). \label{fpi1loop}
\end{eqnarray}
At two-loop order, we have diagrams that involve the fluctuations from the scalar 
$\zeta$, $\vec{\pi}$, and $\phi$ fields, that contribute to $f_{\pi}^{2}$. 
Their expressions are given,  respectively, by

\begin{eqnarray}
if_{\pi,\zeta}^{2}g_{\mu\nu}\delta_{ij}
&=&\frac{iG_{s}N_{c}}{2}
\int \frac{d^{4}p_{1}}{(2\pi)^{4}}\int \frac{d^{4}p_{2}}{(2\pi)^{4}}
\textrm{Tr}_{D} \left[\textrm{Tr}_{f}(\tau_{i}\tau_{j})
  \vphantom{\frac{\slashed{p}_{1}
      +\mathcal{M}}{p_{1}^{2}-\mathcal{M}^{2}+i\epsilon}}
  \right. \nonumber\\ &\times&
  \left. \frac{\slashed{p}_{1}+\mathcal{M}}{p_{1}^{2}
    -\mathcal{M}^{2}+i\epsilon}\gamma_{\mu}\gamma^{5}\frac{\slashed{p}_{1}+
\slashed{q}
    +\mathcal{M}}{(p_{1}+q)^{2}-\mathcal{M}^{2}+i\epsilon}
  \right. \nonumber\\ &\times&\left. \frac{\slashed{p}_{2}+
\slashed{q}+\mathcal{M}}{(p_{2}+q)^{2}
    -\mathcal{M}^{2}+i\epsilon}\gamma_{\nu}\gamma^{5}\frac{\slashed{p}_{2}+
\mathcal{M}}{p_{2}^{2}
    -\mathcal{M}^{2}+i\epsilon}\right],
\nonumber \\
\end{eqnarray}

\begin{eqnarray}
if_{\pi,\vec{\pi}}^{2}g_{\mu\nu}\delta_{ij}
&=&-\frac{iG_{s}N_{c}}{2}\int
\frac{d^{4}p_{1}}{(2\pi)^{4}}\int \frac{d^{4}p_{2}}{(2\pi)^{4}}
\nonumber \\ &\times&
\textrm{Tr}_{D}\left[\textrm{Tr}_{f}(\tau_{k}\tau_{i}\tau_{k}\tau_{j})\gamma^{5}
  \frac{\slashed{p}_{1}+\mathcal{M}}{p_{1}^{2}-\mathcal{M}^{2}+i\epsilon}
  \gamma_{\mu}\gamma^{5}\right. \nonumber\\ &\times&
  \frac{\slashed{p}_{1}+\slashed{q}+\mathcal{M}}{(p_{1}+q)^{2}
    -\mathcal{M}^{2}+i\epsilon}\gamma^{5}\frac{\slashed{p}_{2}+\slashed{q}
    +\mathcal{M}}{(p_{2}+q)^{2}-\mathcal{M}^{2}+i\epsilon} \nonumber\\ &\times&
  \left. \gamma_{\nu}\gamma^{5}\frac{\slashed{p}_{2}
    +\mathcal{M}}{p_{2}^{2}-\mathcal{M}^{2}+i\epsilon}\right],
\end{eqnarray}
and

\begin{eqnarray}
if_{\pi,\phi}^{2}g_{\mu\nu}\delta_{ij}
&=&-iG_{d}(N_{c}-1)\int
\frac{d^{4}p_{1}}{(2\pi)^{4}}\int
\frac{d^{4}p_{2}}{(2\pi)^{4}}\textrm{Tr}_{D} \nonumber\\ &\times&
\left[\textrm{Tr}_{f}(\tau_{2}\tau_{i}^{T}\tau_{2}\tau_{j})\gamma^{5}
  \frac{\slashed{p}_{1}+\mathcal{M}}{p_{1}^{2}-\mathcal{M}^{2}+i\epsilon}
\gamma_{\mu}\gamma^{5}\right. \nonumber\\ &\times&\left. 
\frac{\slashed{p}_{1}+\slashed{q}+\mathcal{M}}{(p_{1}+q)^{2}
    -\mathcal{M}^{2}+i\epsilon}\gamma^{5}\frac{\slashed{p}_{2}+\slashed{q}+
\mathcal{M}}{(p_{2}+q)^{2}
    -\mathcal{M}^{2}+i\epsilon} \right. \nonumber\\ &\times&
  \left. \gamma_{\nu}\gamma^{5}\frac{\slashed{p}_{2}+\mathcal{M}}{p_{2}^{2}
    -\mathcal{M}^{2}+i\epsilon}\right].
\end{eqnarray}
When $q=0$, we have [recalling that
$\textrm{Tr}_{f}(\tau_{i}\tau_{j})=N_{f}\delta_{ij}$,
$\textrm{Tr}_{f}(\tau_{k}\tau_{i}\tau_{k}\tau_{j})=-(n_{\pi}-2)N_{f}\delta_{ij}$
and
$\textrm{Tr}_{f}(\tau_{2}\tau_{i}^{T}\tau_{2}\tau_{j})=-N_{f}\delta_{ij}$]

\begin{eqnarray}
f_{\pi,\zeta}^{2}g_{\mu\nu}&=&\frac{G_{s}N_{c}N_{f}}{2}\int
\frac{d^{4}p_{1}}{(2\pi)^{4}} \int \frac{d^{4}p_{2}}{(2\pi)^{4}}
\nonumber\\ &\times&
\frac{1}{(p_{1}^{2}-\mathcal{M}^{2}+i\epsilon)^{2}(p_{2}^{2}-\mathcal{M}^{2}+i\epsilon)^{2}}
\nonumber \\
&\times& \textrm{Tr}_{D}\left[(\slashed{p}_{1}+\mathcal{M})\gamma_{\mu}\gamma^{5}
  \right. \nonumber\\ &\times&
  \left. (\slashed{p}_{1}+\mathcal{M})(\slashed{p}_{2}+
\mathcal{M})\gamma_{\nu}\gamma^{5}
  (\slashed{p}_{2}+\mathcal{M})\right], \label{fpis0}
\end{eqnarray}

\begin{eqnarray}
f_{\pi,\vec{\pi}}^{2}g_{\mu\nu}&=&\frac{G_{s}N_{c}N_{f}}{2}(n_{\pi}-2)\int
\frac{d^{4}p_{1}}{(2\pi)^{4}}\int \frac{d^{4}p_{2}}{(2\pi)^{4}}
\nonumber\\ &\times&
\frac{1}{(p_{1}^{2}-\mathcal{M}^{2}+i\epsilon)^{2}(p_{2}^{2}-\mathcal{M}^{2}+i\epsilon)^{2}}
\nonumber\\
&\times&
\textrm{Tr}_{D}\left[\gamma^{5}(\slashed{p}_{1}+\mathcal{M})
\gamma_{\mu}\gamma^{5}
  \right. \nonumber\\ &\times&
  \left. (\slashed{p}_{1}+\mathcal{M})\gamma^{5}(\slashed{p}_{2}+\mathcal{M})
  \gamma_{\nu}\gamma^{5}(\slashed{p}_{2}+\mathcal{M})\right], \label{fpips0}
\end{eqnarray}
and

\begin{eqnarray}
f_{\pi,\phi}^{2}g_{\mu\nu} & = &G_{d}(N_{c}-1)N_{f}\int
\frac{d^{4}p_{1}}{(2\pi)^{4}}\int \frac{d^{4}p_{2}}{(2\pi)^{4}}
\nonumber\\ &\times&
\frac{1}{(p_{1}^{2}-\mathcal{M}^{2}+i\epsilon)^{2}(p_{2}^{2}-\mathcal{M}^{2}+i\epsilon)^{2}}
\nonumber\\
&\times&
\textrm{Tr}_{D}\left[\gamma^{5}(\slashed{p}_{1}+\mathcal{M})
\gamma_{\mu}\gamma^{5}
  \right. \nonumber\\ &\times&
  \left. (\slashed{p}_{1}+\mathcal{M})\gamma^{5}(\slashed{p}_{2}+\mathcal{M})
  \gamma_{\nu}\gamma^{5}(\slashed{p}_{2}+\mathcal{M})\right]. \label{fpid0}
\end{eqnarray}
The double integrals involving the  the traces in Eqs.~(\ref{fpips0}) and (\ref{fpid0}) are equivalent,
and we can define, for convenience, the momentum integrals appearing in those
equations as

\begin{eqnarray}
&&F \equiv \int \frac{d^{4}p_{1}}{(2\pi)^{4}}\int
  \frac{d^{4}p_{2}}{(2\pi)^{4}}\frac{1}{(p_{1}^{2}
    -\mathcal{M}^{2}+i\epsilon)^{2}(p_{2}^{2}-\mathcal{M}^{2}+i\epsilon)^{2}}
  \nonumber\\ &&\times
  \textrm{Tr}_{D}\left[(\slashed{p}_{1}+\mathcal{M})\gamma_{\mu}\gamma^{5}
    (\slashed{p}_{1}+\mathcal{M})(\slashed{p}_{2}+\mathcal{M})
\gamma_{\nu}\gamma^{5}
    (\slashed{p}_{2}+\mathcal{M})\right] \nonumber\\ && =
  4\left\{g_{\mu\nu}\left[-I_G^{2}(\mathcal{M})+2\mathcal{M}^{4}I^{2}(0)\right]
  \vphantom{\int \frac{d^{4}p_{1}}{(2\pi)^{4}}}
  \right. \nonumber\\ &&+ \left. 4\int
  \frac{d^{4}p_{1}}{(2\pi)^{4}}\int \frac{d^{4}p_{2}}{(2\pi)^{4}}
  \frac{(p_{1}\cdot
    p_{2})p_{1\mu}p_{2\nu}}{(p_{1}^{2}-\mathcal{M}^{2}+i\epsilon)^{2}(p_{2}^{2}
    -\mathcal{M}^{2}+i\epsilon)^{2}}   \right\}. \nonumber\\ \label{FF}
\end{eqnarray}
The calculations in order to find Eq.~(\ref{FF}) are relatively
laborious but straightforward. The double integral on the right-hand
side in Eq.~(\ref{FF}), which we will denote by $L$, when using dimensional regularization 
and  the relation Eq.~(\ref{DimReg}), becomes

\begin{eqnarray}
L&\equiv&4\int \frac{d^{4}p_{1}}{(2\pi)^{4}}\int
\frac{d^{4}p_{2}}{(2\pi)^{4}} \frac{(p_{1}\cdot
  p_{2})p_{1\mu}p_{2\nu}}{(p_{1}^{2}-\mathcal{M}^{2}+i\epsilon)^{2}(p_{2}^{2}
  -\mathcal{M}^{2}+i\epsilon)^{2}} 
\nonumber\\  
&=&4g^{\alpha\beta}\int
\frac{d^{4}p_{1}}{(2\pi)^{4}}\frac{p_{1\alpha} p_{1\mu}}{(p_{1}^{2}
  -\mathcal{M}^{2}+i\epsilon)^{2}}\int \frac{d^{4}p_{2}}{(2\pi)^{4}}
\frac{p_{2\beta}p_{2\nu}}{(p_{2}^{2}-\mathcal{M}^{2}+i\epsilon)^{2}}
\nonumber\\ &=&4g^{\alpha\beta}\frac{g_{\alpha\mu}}{2}I_{G}(0)
\frac{g_{\beta\nu}}{2}I_{G}(0)
\nonumber\\ &=&g_{\mu\nu}I_{G}^{2}(\mathcal{M}). \label{LL}
\end{eqnarray}
Substituting Eq.~(\ref{LL}) into Eq.~(\ref{FF}), and Eq.~(\ref{FF}) into
Eqs.~(\ref{fpis0}), (\ref{fpips0}), and (\ref{fpid0}), we obtain

\begin{eqnarray}
f_{\pi,\zeta}^{2}&=&4G_{s}N_{c}N_{f}\mathcal{M}^{4}I^{2}(0), \label{fpis}
\end{eqnarray}
\begin{eqnarray}
f_{\pi,\vec{\pi}}^{2}&=&4G_{s}N_{c}N_{f}(n_{\pi}-2)\mathcal{M}^{4}I^{2}(0), \label{fpips}
\end{eqnarray}
\begin{eqnarray}
f_{\pi,\phi}^{2}&=&8G_{d}(N_{c}-1)N_{f}\mathcal{M}^{4}I^{2}(0). \label{fpid}
\end{eqnarray}
The final expression for $f_{\pi}^{2}$ is obtained by summing
Eqs.~(\ref{fpi1loop}),  (\ref{fpis}), (\ref{fpips}), and (\ref{fpid}),
to finally give the result

\begin{eqnarray}
f_{\pi}^{2}&=&-2iN_{c}N_{f}\mathcal{M}^{2}I(0) \nonumber\\ &+&
4N_{c}N_{f}G_{s}\left[(n_{\pi}-1)+\frac{2G_{d}}{G_{s}}\frac{(N_{c}-1)}{N_{c}}\right]
\mathcal{M}^{4}I^{2}(0)
\nonumber\\ &=&-2iN_{c}N_{f}\mathcal{M}^{2}I(0)+4N_{c}N_{f}G_{s}f(G_{d})
\mathcal{M}^{4}I^{2}(0),
\nonumber\\
\end{eqnarray}
where

\begin{eqnarray} \label{deff(Gd)ap}
f(G_{d})&\equiv&(n_{\pi}-1)+\frac{2G_{d}}{G_{s}}\frac{(N_{c}-1)}{N_{c}}.
\end{eqnarray}

%%%%%%%%%%%%%%%%%%%%%%%%%%%%%%%%%%%%%%%%%%%%%%%%%%%%%%%%%%%%%%%%%%%%%%%%%%

\end{document}